\documentclass[useAMS,usenatbib]{mnras}
\usepackage{graphicx}   
\usepackage{amsmath}    
\usepackage{aas_macros} %
\usepackage{subfig}
\usepackage{graphics}

\newcommand{\msun}{\ensuremath{\textup{ M}_{\odot}}}                   
\newcommand{\lsun}{\ensuremath{\textup{ L}_{\odot}}}                   
\newcommand{\mjup}{\ensuremath{\textup{ M}_{\textsc{j}}}}              
\newcommand{\dif}{\mathrm{d}}                                          
\renewcommand{\vec}[1]{\boldsymbol{#1}}                                
\newcommand{\scell}[2][c]{\begin{tabular}[#1]{@{}c@{}}#2\end{tabular}} 

\title[Disc fragmentation with radiative feedback]{The effect of radiative feedback on disc fragmentation}
\author[Anthony Mercer \& Dimitris Stamatellos]
{Anthony~Mercer$^{1,}$\thanks{E-mail: \texttt{apmercer@uclan.ac.uk}} \& Dimitris~Stamatellos$^1$ 
\\ 
$^1$Jeremiah Horrocks Institute for Mathematics, Physics \& Astronomy, University of Central Lancashire, Preston, PR1 2HE, UK}

\date{Accepted XXX. Received YYY; in original form ZZZ}

\pubyear{2016}

\begin{document}
\label{firstpage}
\pagerange{\pageref{firstpage}--\pageref{lastpage}}


\maketitle

\begin{abstract}
	Protostellar discs may become massive enough  to fragment producing secondary low-mass objects: planets, brown dwarfs and low-mass stars. We study the effect of radiative feedback from such newly-formed secondary objects using radiative hydrodynamic simulations. We compare the results of simulations without any radiative feedback from secondary objects with those where two types of radiative feedback are considered: (i) continuous, and (ii) episodic.  We find that: (i) continuous radiative feedback stabilizes the disc and suppresses further fragmentation, reducing the number secondary objects formed; (ii) episodic feedback from secondary objects heats and stabilises the disc when the outburst occurs, but shortly after the outburst stops, the disc becomes unstable and fragments again. However, fewer secondary objects are formed compared to the the case without radiative feedback. We also find that the mass growth of secondary objects is mildly suppressed due to the effect of their radiative feedback. However, their mass growth also depends on where they form in the disc and on their subsequent interactions, such that their final masses are not drastically different from the case without radiative feedback. We find that the masses of secondary objects formed by disc fragmentation are from a few M$_{\rm J}$ to a few 0.1~M$_{\sun}$. Planets formed by fragmentation tend to be ejected from the disc. We conclude that planetary-mass objects on wide orbits (wide-orbit planets) are unlikely to form by disc fragmentation. Nevertheless, disc fragmentation may be a significant source of free-floating planets and brown dwarfs.
\end{abstract}

\begin{keywords}
    methods: numerical - planetary systems: planet-disc interactions, protoplanetary discs - stars: brown dwarfs
\end{keywords}

\section{Introduction} 
\label{sec:introduction}

    Stars form from the gravitational collapse of dense cores within turbulent molecular clouds. Due to turbulence and/or the initial rotation of these cores, newly formed stars are attended by protostellar discs of gas and dust \citep[e.g.][]{Terebey:1984a,Attwood:2009a}. The idea that our Solar System formed from a protostellar disc has been discussed since the 18th century \citep{Laplace:1796a}. We now know that discs play a fundamental role in the formation of stars and planets. They provide young stars with the majority of their mass through accretion and a suitable dynamical and chemical environment for the formation of planets \citep[e.g.][]{Lissauer:1993a}.

    Mass accretion onto young stars  increases their luminosity due to gravitational energy being converted into heat on the accretion shock around the surface of a star. The accretion is typically considered to be continuous \citep{Krumholz:2006a,Bate:2009a,Offner:2009a,Krumholz:2010b}. However, consider a star which has just evolved out of the Class 0 phase, which will ultimately have a mass of $1 \msun$. It has an age of $\sim 10^{5}$ yr and has accumulated half of its final mass. This yields a mean mass accretion rate of $\sim 5 \times 10^{-6} \msun \textup{yr}^{-1}$ and a mean luminosity of $\sim 25 \lsun$. Observational studies show that solar-like protostars have much lower luminosities \citep[e.g.][]{Kenyon:1990a,Evans:2009b, Enoch:2009a}. This is the so called {\it luminosity problem}. It may be circumvented if accretion onto protostars is not continuous, but rather episodic, happening in short bursts \citep[e.g.][]{Dunham:2010a,Dunham:2012a,Audard:2014a}. 

    FU Ori objects provide evidence of episodic accretion. These objects exhibit sudden luminosity increases on the order of $\sim5$~mag and estimated accretion rates of $> 10^{-4} \msun \textup{yr}^{-1}$ which last from a few tens of years to a few centuries \citep{Herbig:1977a, Hartmann:1996a,Greene:2008a,Peneva:2010a,Green:2011a}.

    Episodic accretion maybe due to gravitational instabilities \citep{Vorobyov:2005a,Machida:2011a,Vorobyov:2015a,Liu:2016b}, thermal instabilities in the inner disc region \citep{Hartmann:1985a, Lin:1985a, Bell:1994a}, or due to gravitational interactions in a binary system \citep{Bonnell:1992a,Forgan:2010a}. It has also been suggested that they may be due to the combined effect of gravitational instabilities operating in the outer disc region transferring matter inwards and the magneto-rotational instability (MRI) operating episodically in the inner disc region delivering matter onto the young protostar \citep{Armitage:2001a,Zhu:2007a,Zhu:2009a,Zhu:2009b,Zhu:2010a}.

    It is expected that radiative feedback from young protostars will affect the dynamical and thermal evolution of their parent cloud and their discs \citep{Stamatellos:2011a,Stamatellos:2012a,Lomax:2014a,Guszejnov:2016a}.
    A significant fraction of low-mass stars and brown dwarfs may form by fragmentation in gravitationally unstable discs \citep{Whitworth:2006a, Stamatellos:2007c, Stamatellos:2009a,Kratter:2010a,Zhu:2012a,Lomax:2014a,Kratter:2016a,Vorobyov:2013a}. Protostellar discs fragment if two conditions are met: (i) They are gravitationally unstable i.e. 
    \begin{equation}
        Q \equiv \frac{\kappa c_{s}}{\pi G \Sigma} < \beta,
    \label{eqn:introduction:toomre}
    \end{equation}
    where $Q$ is the Toomre parameter \citep{Toomre:1964a}, $\kappa$ is the epicyclic frequency, $c_{s}$ is the local sound speed and $\Sigma$ is the disc surface density. The value of $\beta$ is on the order of unity and it is dependant on the assumed geometry of the disc and the equation of state used: for a razor-thin disc, $\beta = 1$; in a 3D disc $\beta = 1.4$ \citep[see][and references therein]{Durisen:2007a}. (ii)~They cool sufficiently fast, i.e. $t_{\textsc{cool}}<(0.5-2)t_{\textsc{orb}}$ where $t_{\textsc{orb}}$ is the local orbital period \citep{Gammie:2001a,Johnson:2003a, Rice:2003a, Rice:2005a}. In the last few years, the validity of second criterion has been scrutinized, and it has been suggested that fragmentation may happen for even slower cooling rates \citep{Meru:2011a,Paardekooper:2011a,Lodato:2011a, Rice:2012a, Tsukamoto:2015a}. However, irrespective of the detailed criteria, there has been significant observational evidence that disc fragmentation does occur \citep{Tobin:2013b, Tobin:2016a, Dupuy:2016a}.

    Theoretical work and numerical simulations suggest that the conditions for disc fragmentation are met in the outer disc regions ($>70-100$~AU) \citep[e.g.][]{Whitworth:2006a,Stamatellos:2011d,Stamatellos:2009d,Stamatellos:2008a,Boley:2009a}. Most of the objects formed by disc fragmentation are brown dwarfs, though low-mass stars and planets may also form \citep{Stamatellos:2009a, Zhu:2012a,Vorobyov:2013a}. Fragments that form in gravitationally unstable discs start off with a mass that is determined by the opacity limit for fragmentation, i.e. with a few ${\rm M}_{\rm J}$, where M$_{\rm J}$ is the mass of Jupiter \citep{Low:1976a,Rees:1976a,Boss:1988a,Whitworth:2006a,Boley:2010b,Forgan:2011b,Rogers:2012a}. However, they quickly accrete mass to become brown dwarfs or even low-mass stars \citep{Stamatellos:2009a,Kratter:2010b,Zhu:2012a, Stamatellos:2015a}. A few of the fragments remain in the planetary-mass regime ($M<13~{\rm M}_{\rm J}$) but these are typically ejected from the disc \citep{Li:2015b,Li:2016a} becoming free-floating planets \citep[e.g.][]{Osorio:2000a, Kellogg:2016a}.

    These low-mass objects that form by disc fragmentation have properties that are similar to the properties of objects forming from the collapse of isolated low-mass pre-(sub)stellar cloud cores. They are expected to be attended by discs \citep{Stamatellos:2009a,Liu:2015a,Sallum:2015a}, and they may also launch jets perpendicular to the disc axis \citep{Machida:2006a, Gressel:2015a}. \cite{Stamatellos:2015b} suggest that discs around low-mass objects (brown dwarfs and planets) that form by disc fragmentation are more massive from what would be expected if they were formed in collapsing low-mass pre-(sub)stellar cloud cores, which is consistent with recent observations of brown dwarf discs in Upper Sco OB1 and Ophiuchus \citep{van-der-Plas:2016a}. 

    It is therefore reasonable to assume that low-mass objects that form by disc fragmentation may also exhibit radiative feedback due to accretion of material  from their individual discs onto their surfaces. The effect of radiative feedback due to  accretion onto low-mass objects such as planets and brown dwarfs has been ignored by previous studies of disc fragmentation. Recent simulations of the evolution of giant proto-planets in self-gravitating discs \citep{Stamatellos:2015a} have shown that radiative feedback from giant planets may reduce gas accretion  and hence suppress their mass growth. They found that when radiative feedback is included  the fragment's final mass is within the planetary-mass regime \citep[see also][]{Nayakshin:2013a}.

    The goal of this paper is to examine how radiative feedback from objects that form by disc fragmentation influences the properties of these objects and whether subsequent fragmentation in the disc is affected. More specifically we  investigate whether radiative feedback from objects forming in the disc (hereafter referred to as {\it secondary objects}) suppresses their mass growth, increasing the possibility that these objects will end up as planets rather than brown dwarfs of more massive objects, in contrast with what previous studies suggest \citep[e.g.][]{Stamatellos:2009a,Kratter:2010b}.

    We construct numerical experiments to examine three cases of radiative feedback from secondary objects: {\bf (i)}~No radiative feedback: gas is accreted onto the objects but no energy is fed back into the disc due to this process. {\bf (ii)}~Continuous radiative feedback:  gas is accreted continuously onto the object and the accretion energy is continuously fed back into the disc. {\bf (iii)}~Episodic radiative feedback: We assume that low mass secondary objects exhibit episodic outbursts just like their higher-mass counterparts do. Gas accumulates into the region close to the object (within $\sim 1$~AU) and when the conditions are right it accretes onto the object (see Section~\ref{sec:numerical_method} for details). Gas accretion onto secondary objects is episodic, resulting in episodic radiative feedback.
 
    In Section~\ref{sec:numerical_method}, we provide the computational framework of this work including the episodic accretion/feedback model we adopt. In Section~\ref{sec:initial_conditions} we discuss the initial conditions of the simulations. We present the results of the effect of radiative feedback on the evolution of discs and on the properties of the objects form by disc fragmentation in Section~\ref{sec:disc_fragmentation_and_the_effect_of_radiative_feedback}. Our results are summarised in Section~\ref{sec:conclusions}.


\section{Numerical Method} 
\label{sec:numerical_method}
 
    We use the smoothed-particle hydrodynamics (SPH) code \textsc{seren} \citep{hubber:2011a,hubber:2011b} to simulate gravitationally unstable protostellar discs.  Discs are represented by a large number of SPH particles. To avoid small timesteps at a density  of $\rho_{\rm sink}=10^{-9} \textup{ g cm}^{-3}$ a particle is replaced by a sink \citep{Bate:1995a} that represents a bound object (star, brown dwarf or planet, depending on its mass). Sinks interact with the rest of the disc both gravitationally and radiatively (in the cases where radiative feedback is included). Gas particles which pass within $R_{\rm sink}=1$~AU and are gravitationally bound to a sink are accreted onto it.

    The heating and cooling of gas is performed using the radiative transfer technique ascribed to \cite{Stamatellos:2007b}, where the density and the gravitational potential of a gas particle are used to estimate a column density through which cooling/heating happens, and along with the local opacity, are used to estimate an optical depth for each particle. This can be used to determine the heating and cooling of the particle and incorporates effects from the rotational and vibrational degrees of freedom of $\textup{H}_{2}$, the dissociation of $\textup{H}_{2}$, ice melting, dust sublimation, bound-free, free-free and electron scattering interactions. The equation of state used and the effect of each assumed constituent are described in detail in \textsection 3 of \cite{Stamatellos:2007b}.
    The radiative heating/cooling rate of a particle $i$ is
    \begin{equation}
        \frac{\dif u_{i}}{\dif t} = \frac{4 \sigma_{\textsc{sb}} \left( T_{\textsc{bgr}}^{4} - T_{i}^{4} \right)}{\bar{\Sigma}_{i}^{2} \bar{\kappa}_{\textsc{r}}\left(\rho_{i}, T_{i} \right) + \kappa_{\textsc{p}}^{-1}\left(\rho_{i}, T_{i} \right)},
    \label{eqn:heatingRate}
    \end{equation}
    where $\sigma_{\textsc{sb}}$ is the Stefan-Boltzmann constant, $T_{\textsc{bgr}}$ is the pseudo-background temperature below which particles cannot cool radiatively, $\bar{\Sigma}_{i}$ is mass-weighted mean column density of the particle, and $\bar{\kappa}_{\textsc{r}}$ and $\kappa_{\textsc{p}}$ are the Rosseland- and Planck-mean opacities, respectively. 
 
     Once most of the gas in the disc has dissipated (accreted onto the central star and onto the secondary objects formed in the disc; $t=10$~kyr), we utilise an N-body integrator with a 4th-order Hermite integration scheme \citep{Makino:1992a}, to follow the evolution of the objects present at the end of each hydrodynamic simulation up to $200 \textup{ kyr}$. We use a strict timestep criterion so that energy is conserved to better than one part in $10^8$ \citep{Hubber:2005a}. This allows us to determine the ultimate fate of these objects: will they remain bound to central star or be ejected from the system? It is noted that at this phase we ignore gravitational and dissipative interactions due to gas within the disc.   
  
    \subsection{Radiative feedback from sinks} 
    \label{sub:radiative_feedback_from_sinks_stars}
        Sinks that represent stars, brown dwarfs and planets in the simulations interact both gravitationally and radiatively with the disc. In the optically thin limit, the temperature that the dust/gas will attain at  a distance $\left| \vec{r} - \vec{r}_{n}\right|$ from a radiative object $n$ is    
\begin{equation}
        T_n\left(\vec{r} \right) =  \left( \frac{L_{n}}{16 \pi \sigma_{\textsc{sb}}} \right)^{1/4}
          \left( \left| \vec{r} - \vec{r}_{n}\right| \right)^{-1/2}\,.
    \label{eqn:Tbgr2}
    \end{equation}
In the optically thick limit, considering a geometrically thin, passive disc \cite[e.g.][]{Kenyon:1987a} the temperature is
      \begin{equation}
        T_n\left(\vec{r} \right) =  \left( \frac{L_{n} R_{n}}{4 \pi \sigma_{\textsc{sb}} } \right)^{1/4}
         \left( \left| \vec{r} - \vec{r}_{n}\right| \right)^{-3/4}\,.
    \label{eqn:Tbgr3}
    \end{equation}
Therefore, the temperature drops faster with the distance from the radiative object in the optically thick case ($q=3/4$ vs $q=1/2$, respectively). However, in the case of a flared disc the temperature drop is less steep, approaching the $q=1/2$ value. This is because a flared disc intercepts a higher fraction of the star's radiation \citep[e.g.][]{Kenyon:1987a, Chiang:1997a}. This lower value for $q$ is also consistent with disc observations \cite[e.g.][]{Andrews:2009b}. 

	Customarily, the optically thin case is used in analytic and computational studies of protostellar disc evolution \citep[e.g.][]{Matzner:2005a,Kratter:2006a,Stamatellos:2007b,Offner:2009a,Stamatellos:2009d,Stamatellos:2011a,Zhu:2012a,Lomax:2014a,Vorobyov:2015a,Dong:2016a,Kratter:2016a}, albeit with a scaled down stellar luminosity  (by a factor of $\sim 0.1$) so as to match detailed radiative transfer calculations \citep[see][]{Matzner:2005a}. In either case, the temperature at a given distance from a radiative source depends on the luminosity of the source. The luminosity of young stellar and substellar objects is mostly due to accretion of material onto their surfaces.

 	In the simulations presented here we assume a time-independent contribution from the central star in the optically thick regime, and a time-dependent contribution in the optically thin regime from the secondary objects that form self-consistently in the simulations. We describe each one in detail in the following sections. We note that these contributions only account for disc heating due to radiation released on the surfaces of bound objects; energy release in the disc midplane due to accretion is taken into account self-consistently within the hydrodynamic simulation. This approach ignores the case in which the density of the gas within the Hill radius of a secondary object is high, shielding the rest of the disc from the effect of heating. However, such a phase would be short-lived as gas is accreted onto the secondary object.

        \subsubsection{Radiative feedback from the central star} 
        \label{subsub:radiative_feedback_from_the_central_star}
        We assume that the radiative feedback from the central star is constant with time, and independent of the accretion rate onto it. This is done because the central star is part of the initial conditions and does not form self-consistently in the simulations. Therefore the accretion rate onto it may not be properly determined. Additionally, by choosing a relatively steep temperature profile we  minimise the role of the central star in stabilising the disc, and focus on the radiative effect from the secondary objects forming in the disc.
 
        The pseudo-background temperature due to the central star is set to 
            \begin{equation}
                T_{\textsc{bgr}}^{\star}(R) = \left[ T_{0}^{2} \left( \frac{R^{2} + R_{0}^{2}}{\textup{AU}^{2}} \right)^{-3/4} + T_{\infty}^{2} \right]^{1/2}.
            \label{eqn:temperatureProfile}
            \end{equation}
            $R$ is the distance from the star on the disc midplane, $R_{0} = 0.25$ AU is a smoothing radius which prevents non-physical values when $R \rightarrow 0$, $T_{0} = 250$ K is the temperature at a distance of 1 AU from the central star, and $T_{\infty} = 10$ K is the temperature at large distances from the star. The above equation is chosen purely on numerical grounds to reproduce the required properties of the temperature profile.
            
        \subsubsection{Radiative feedback from secondary objects} 
        \label{subsub:radiative_feedback_from_secondary_objects}

The radiative feedback from secondary objects depends on the accretion rate of gas onto them, and it is therefore time-dependent.
 The pseudo-background temperature  due to  radiative secondary objects in the disc is set to
     \begin{equation}
        T_{\textsc{bgr}}^{4}\left(\vec{r} \right) = \left(10 \textup{ K} \right)^{4} + \sum_{n} \left( \frac{L_{n}}{16 \pi \sigma_{\textsc{sb}} \left| \vec{r} - \vec{r}_{n}\right|^{2}} \right),
    \label{eqn:Tbgr}
    \end{equation}
    where $L_{n}$ and $\vec{r}_{n}$ are the luminosity and position of a radiative object $n$ \citep{Stamatellos:2011a,Stamatellos:2012a,Stamatellos:2015a}. The luminosity of a radiative secondary object $n$ is set to
    \begin{equation}
            L_{n} = L_{\rm NB} + \frac{fGM_{n}\dot{M_{n}}}{R_{\textup{acc}}}.
    \label{eqn:massAccretionLuminosity}
    \end{equation}
    The first term on the right hand side of the equation describes the luminosity of the object from nuclear burning which is set equal  to $\left({M_{n}}/{\msun}\right)^{3}\lsun$ for stellar objects ($M>0.08\msun$) and 0 for substellar objects. The second term represents the accretion luminosity. We let $f = 0.75$ be the fraction of accretion energy that is radiated away at the photosphere of the object \citep{Offner:2010d}. $R_{\textup{acc}}$ is the accretion radius, set to $R_{\textup{acc}} = 3 \textup{ R}_{\odot}$ \citep{Palla:1993a}. The choice of the accretion radius does not qualitatively affect the results presented in this paper.

            We consider three cases of radiative feedback from secondary objects forming in the disc by fragmentation: (i) no radiative feedback, (ii) continuous radiative feedback, and (iii) episodic radiative feedback. In the case of no radiative feedback, objects accrete gas but the accretion energy deposited on their surfaces is not fed back into the disc. In the continuous radiative feedback case, gas accretes onto the object releasing energy that is fed back into the disc through the pseudo-background temperature set by Equations~(\ref{eqn:Tbgr})-(\ref{eqn:massAccretionLuminosity}). In the episodic radiative feedback case, mass accretes in periodic bursts resulting in episodic energy release. 
            
            The episodic accretion model that we use is described in detail in \cite{Stamatellos:2011a} and \cite{Stamatellos:2012a}. Gravitational instabilities cannot develop within the inner regions of a disc ($\sim$ a few AU) around a secondary object due to high temperatures. Therefore, there is no mechanism to transport angular momentum outwards for the gas to accrete onto the object, and mass accumulates in the inner disc region. The accumulation of gas increases the density and temperature. When the temperature is sufficiently high to ionise the gas, the magnetorotational instability (MRI) is activated, and gas starts flowing onto the secondary object. As with gravitational instability, angular momentum is transported outwards and matter flows inward. When the mass in the inner accretion disc is depleted, the MRI ceases, and mass once again begins to accumulate within the inner disc region.

            As the hydrodynamic simulations do not have the resolution to capture the details of the inner accretion disc around each secondary object, \cite{Stamatellos:2011a} developed a sub-grid model to capture the effect of MRI, utilizing the  time-dependant episodic accretion model ascribed to \cite{Zhu:2010a}. Each secondary sink is notionally split into two components, the {\it object} and the {\it  inner accretion disc} (IAD) such that,
            \begin{equation}
                M_{\textup{sink}} = M_{\star} + M_{\textsc{iad}},
            \label{eqn:EAMsinkMass}
            \end{equation}
            where $M_{\star}$ is the mass of the object and $M_{\textsc{iad}}$ is the mass of its inner accretion disc. The accretion rate onto the object, $\dot{M_{\star}}$, is assumed to have two components: a small continuous accretion $\dot{M}_{\textsc{con}}$, and the accretion due to the MRI, $\dot{M}_{\textsc{mri}}$. The total accretion rate is therefore
            \begin{equation}
                \dot{M}_{\star} = \dot{M}_{\textsc{con}} + \dot{M}_{\textsc{mri}}.
            \label{eqn:EAMsinkAccretion}
            \end{equation}
            The material only couples to the magnetic field when it becomes ionised. The temperature at which this occurs is set to $T_{\textsc{mri}} \sim 1400$ K. \cite{Zhu:2010a} estimate that the accretion rate during an episode and the duration of an episode are
            \begin{equation}
                \dot{M}_{\textsc{mri}} \sim 5 \times 10^{-4} \msun \textup{yr}^{-1} \left(\frac{\alpha_{\textsc{mri}}}{0.1}\right),
            \label{eqn:EAMepisodeAccretion}
            \end{equation}
            and
            \begin{equation}
                \Delta t_{\textsc{mri}} \sim 0.25 \textup{ kyr} \left( \frac{\alpha_{\textsc{mri}}}{0.1} \right)^{-1} \left( \frac{M_{\star}}{0.2 \msun} \right)^{2/3} \left( \frac{\dot{M}_{\textsc{iad}}}{10^{-5} \msun \textup{yr}^{-1}} \right)^{1/9}\,,
            \label{eqn:EAMepisodeDuration}
            \end{equation}
            respectively. $\dot{M}_{\textsc{iad}}$ is the mass accretion rate which flows onto the inner accretion disc, i.e. the accretion rate onto the sink. $\alpha_{\textsc{mri}}$ is the MRI viscosity $\alpha$-prescription parameter \citep{Shakura:1973a}. The MRI is assumed to occur when sufficient mass has been accumulated within the inner accretion disc such that
            \begin{equation}
                M_{\textsc{iad}} > M_{\textsc{mri}} \sim \dot{M}_{\textsc{mri}} \Delta t_{\textsc{mri}}.
            \label{eqn:EAMmriCondition}
            \end{equation}
            Substituting in Equations (\ref{eqn:EAMepisodeAccretion}) and (\ref{eqn:EAMepisodeDuration}) yields
            \begin{equation}
                M_{\textsc{iad}} > 0.13 \msun \left( \frac{M_{\star}}{0.2 \msun} \right)^{2/3} \left( \frac{\dot{M}_{\textsc{iad}} }{10^{-5} \msun \textup{yr}^{-1}} \right)^{1/9}.
            \label{eqn:EAMmriConditionDetail}
            \end{equation}
            Observations of FU Orionis stars \citep[see e.g.][]{Hartmann:1996a}, show that the the accretion rate during an outburst episode drops exponentially. We therefore set for the accretion rate onto the central object
            \begin{equation}
                \dot{M}_{\textsc{mri}} = 1.58 \frac{M_{\textsc{mri}}}{\Delta t_{\textsc{mri}}} \exp \left\{ -\frac{t - t _{0}}{\Delta t_{\textsc{mri}}} \right\}, \hspace{1cm} t_{0} < t < t_{0} + \Delta t_{\textsc{mri}}.
            \label{eqn:EAMmriAccretion}
            \end{equation}
            $t_{0}$ and $t_{0} + \Delta t_{\textsc{mri}}$ are the temporal bounds of the accretion episode. The factor of $1.58 = e / (e - 1)$ is included to allow all of the mass in the IAD to be accreted onto the object within $\Delta t_{\textsc{mri}}$. The accumulation of mass into the inner accretion disc occurs on a timescale
            \begin{equation}
                \Delta t_{\textsc{acc}} \sim \frac{M_{\textsc{mri}}}{\dot{M}_{\textsc{iad}}}.
            \label{eqn:EAMmriAccumulationTime}
            \end{equation}
            Using Equation (\ref{eqn:EAMmriConditionDetail}) gives
            \begin{equation}
                \Delta t_{\textsc{acc}} \simeq 13 \textup{ kyr} \left( \frac{M_{\star}}{0.2 \msun} \right)^{2/3} \left( \frac{\dot{M}_{\textsc{iad}}}{10^{-5} \msun \textup{yr}^{-1}} \right)^{-8/9}.
            \label{eqn:EAMmriAccumulationTimeDetail}
            \end{equation}
            Comparing this with Equation (\ref{eqn:EAMepisodeDuration}) shows that the period when mass is being accumulated into the inner accretion disc is much longer than the accretion episodes.

            The free variables in this model are $\dot{M}_{\textsc{con}}$ and $\alpha_{\textsc{mri}}$. Increasing $\alpha_{\textsc{mri}}$ yields shorter but more intense accretion episodes. Note that Equations (\ref{eqn:EAMmriConditionDetail}) and (\ref{eqn:EAMmriAccumulationTimeDetail}) are independent of $\alpha_{\textsc{mri}}$. The uncertainty on $\alpha_{\textsc{mri}}$, which lies in the range $0.01-0.4$ \citep{King:2007a}, is therefore not reflected in the mass accreted in an episode nor the time interval between successive episodes.


\section{Initial Conditions} 
    \label{sec:initial_conditions}
    We study the evolution of a $0.3$-M$_{\sun}$  gravitationally unstable protostellar disc around a $0.7$-M$_{\sun}$ star.  The surface density and temperature profiles of the disc are set to  $\Sigma \propto R^{-p}$ and $T \propto R^{-q}$, respectively. The surface density power index $p$ is thought to lie between 1 and $3/2$ from semi-analytical studies of cloud collapse and disc creation \citep{Lin:1990a,Tsukamoto:2015a}. The temperature power index $q$ has been observed to lie in the range from $0.35$ to $0.8$ from studies of pre-main sequence stars \citep{Andrews:2009b}. Here, we adopt $p = 1$ and a relatively high value of $q = 0.75$, in order to minimize the role of the central star in stabilizing the disc and focus on the radiative effect from the secondary objects forming in the disc.

    The disc extends from an inner radius $R_{\textup{in}} = 1$ AU to an outer radius $R_{\textup{out}} = 100$~AU. The surface density profile we use is
    \begin{equation}
        \Sigma(R) = \Sigma_{0} \left( \frac{R_{0}^{2}}{R_{0}^{2} + R^{2}} \right)^{1/2}\,,
    \label{eqn:columnDensityProfile}
    \end{equation}
    where $\Sigma_{0} = 1.7 \times 10^{4} \textup{ g cm}^{-2}$ is the surface density at $R = 0$. 
    The initial disc temperature profile is set using Equation \ref{eqn:temperatureProfile}, i.e. initially $T(R)=T_{\textsc{BGR}}^{\star}(R)$. We use  $N = 10^{6}$ SPH particles to represent the disc. These are distributed using random numbers between $R_{\textup{in}}$ and $R_{\textup{out}}$ so as to reproduce the disc density profile. The values we use for all the aforementioned parameters are listed in Table \ref{tab:simParams}.

\begin{figure}
        \begin{center}
            \includegraphics[width = 0.5\textwidth, trim = 0cm 0cm 0cm 0cm, clip=true]{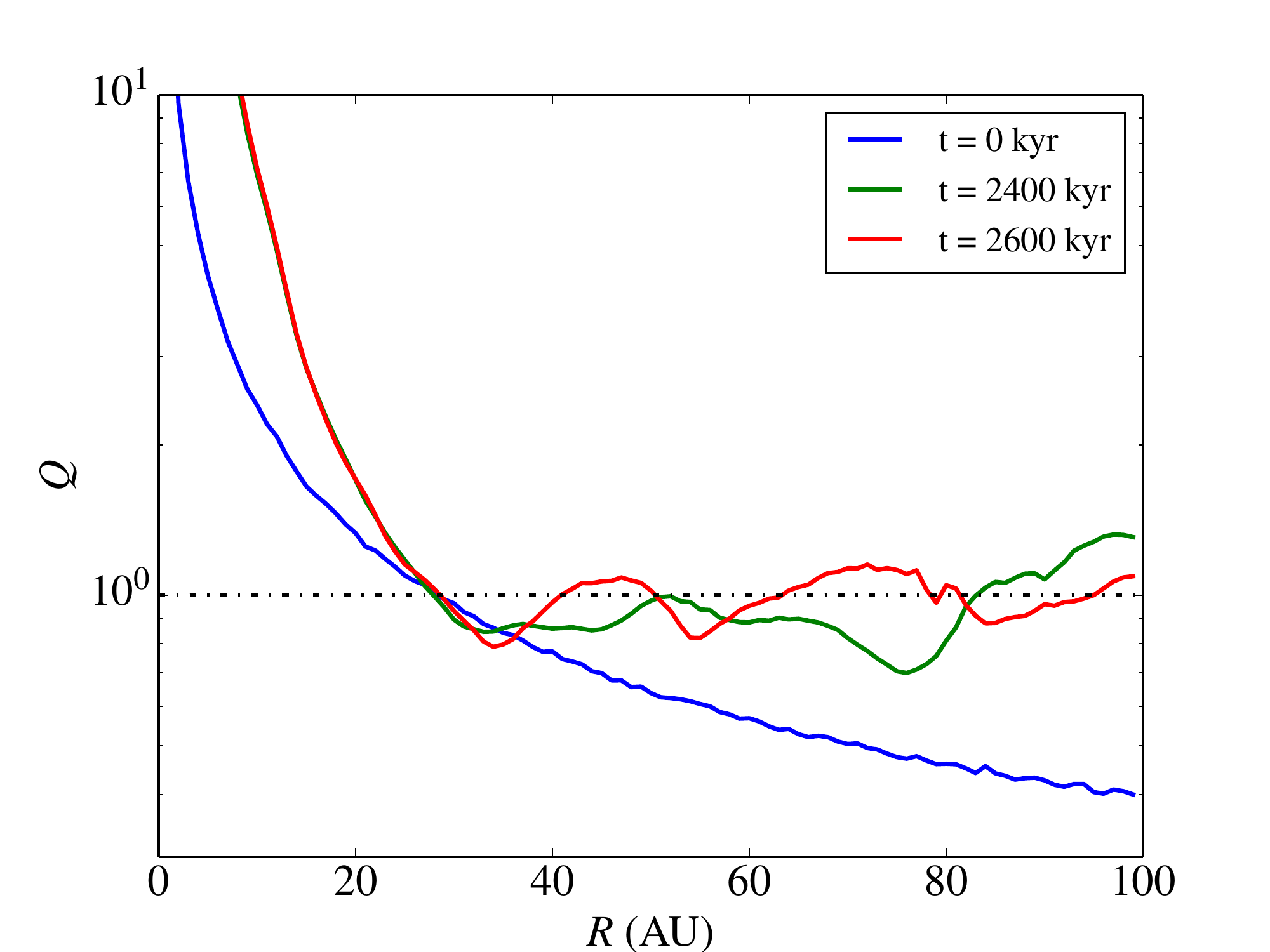}
            \caption{Azimuthally averaged Toomre parameter $Q$ for a disc with the initial conditions listed in Table \ref{tab:simParams}, and at later times (as marked on the graph), before the disc fragments.}
            \label{fig:initialToomre}
        \end{center}
    \end{figure}

The disc is initially massive enough that it is gravitationally unstable ($Q<1$) beyond $\sim 30 \textup{ AU}$ (see Figure \ref{fig:initialToomre}). We have chosen such a profile to ensure that the disc will fragment, so as to study the effect of radiative feedback from secondary objects on subsequent fragmentation. The initial Toomre parameter  reaches very low values at the outer edge of the disc which is unrealistic. When a disc forms around a young protostar its mass increases by infalling material from the protostellar envelope. That progressively reduces $Q$ to just below $\sim1$ and the disc may  then fragment \cite[e.g.][]{Stamatellos:2011a}. In the simulations that we present here, the initial low $Q$ value results in high effective viscosity, so that the disc attains a nearly uniform $Q\sim1$ (see Figure \ref{fig:initialToomre}, red \& green lines) within a few outer orbital periods. This is similar to what it would be expected for a disc forming in a collapsing cloud.
    
  We compute the gravitational acceleration for every disc particle using a spatial octal-tree \citep{Barnes:1986a}. The velocity of a particle $i$ in the $x-y$ plane are set  using $v_{xy, i} = \sqrt{R_{i} \left| g_{xy, i} \right|}$, where $R_{i}$ and $g_{xy, i}$ are the radius  and  the gravitational acceleration of the particle on the disc midplane, respectively. We assume that there are no initial motions perpendicular to the disc midplane.
    
    The number of SPH particles used to represent the disc $\left( 10^{6} \right)$ ensures that gravitational fragmentation can be properly resolved. The minimum resolvable mass for a $0.3 \msun$ disc comprising $10^{6}$ particles is $3.14 \times 10^{-4} \mjup$. \cite{Bate:1997a} argue that the Jeans mass must be resolved by $2 \times N_{\textup{neigh}}$ and \cite{Nelson:2006a} conclude that the Toomre mass must be resolved by $6 \times N_{\textup{neigh}}$. The simulations performed by \cite{Stamatellos:2009d} with a $0.7 \msun$ disc and $1.5 \times 10^{5}$ particles find a minimum Jeans mass of $\sim 2 \mjup$ and a minimum Toomre mass of $\sim 2.5 \mjup$. If we take $2 \mjup$ as a lower resolution limit, then this corresponds to $\sim 128 \times N_{\textup{neigh}}$ i.e. the disc is sufficiently resolved. The vertical structure of our disc is also adequately resolved, since we use $\sim 7$ times more particles than the simulations of \cite{Stamatellos:2009d} where the disc scale-height is resolved by a factor of more than $3-5$ smoothing lengths.
    
    \begin{center}
        \begin{table}
        \centering
            \caption{The initial disc parameters. The disc is gravitationally unstable, as determined by the Toomre criterion.}
            \begin{tabular}{l r}
                \hline
                \hline

                Disc Parameter & Value \\
                \hline
                \hline
                $N$                  & $10^{6}$ \\
                $M_{\textup{disc}}$  & $0.3 \msun$ \\
                $M_{\star}$          & $0.7 \msun$ \\
                $R_{\textup{in}}$    & $1 \textup{ AU}$ \\
                $R_{\textup{out}}$   & $100 \textup{ AU}$ \\
                $R_{0}$              & $0.25 \textup{ AU}$ \\
                $T_{0}$              & $250 \textup{ K}$ \\
                $T_{\infty}$         & $10 \textup{ K}$ \\
                $p$                  & 1 \\
                $q$                  & 0.75 \\
                \hline
            \end{tabular}
            \label{tab:simParams}
        \end{table}
    \end{center}

\section{Disc fragmentation and the Effect of Radiative Feedback} 
\label{sec:disc_fragmentation_and_the_effect_of_radiative_feedback}

\begin{figure*}
            \begin{centering}
            \subfloat{\includegraphics[width = 0.75\textwidth, trim = 0cm 6cm 0cm 0cm, clip=true]{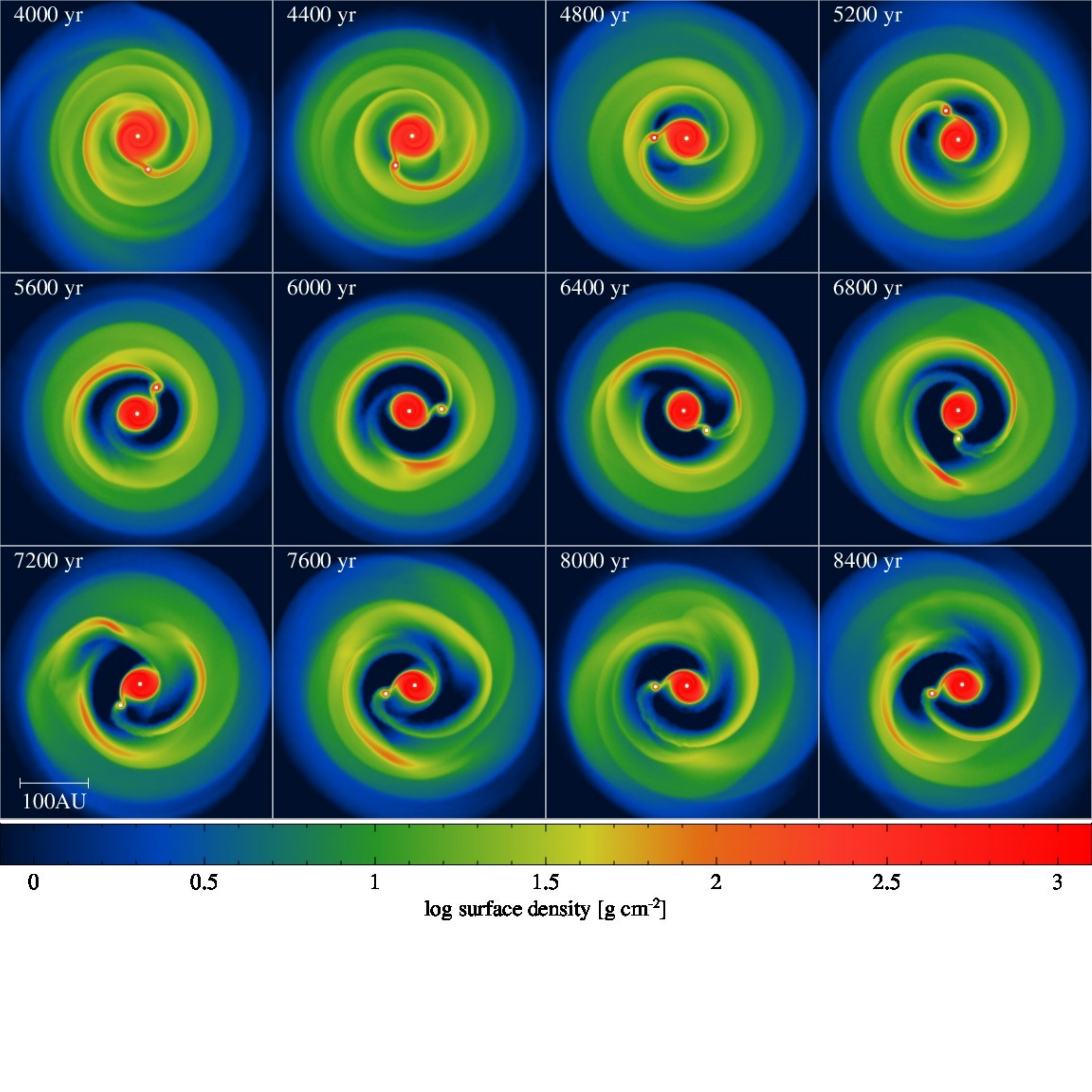}} \\
            \subfloat{\includegraphics[width = 0.75\textwidth, trim = 0cm 6cm 0cm 0cm, clip=true]{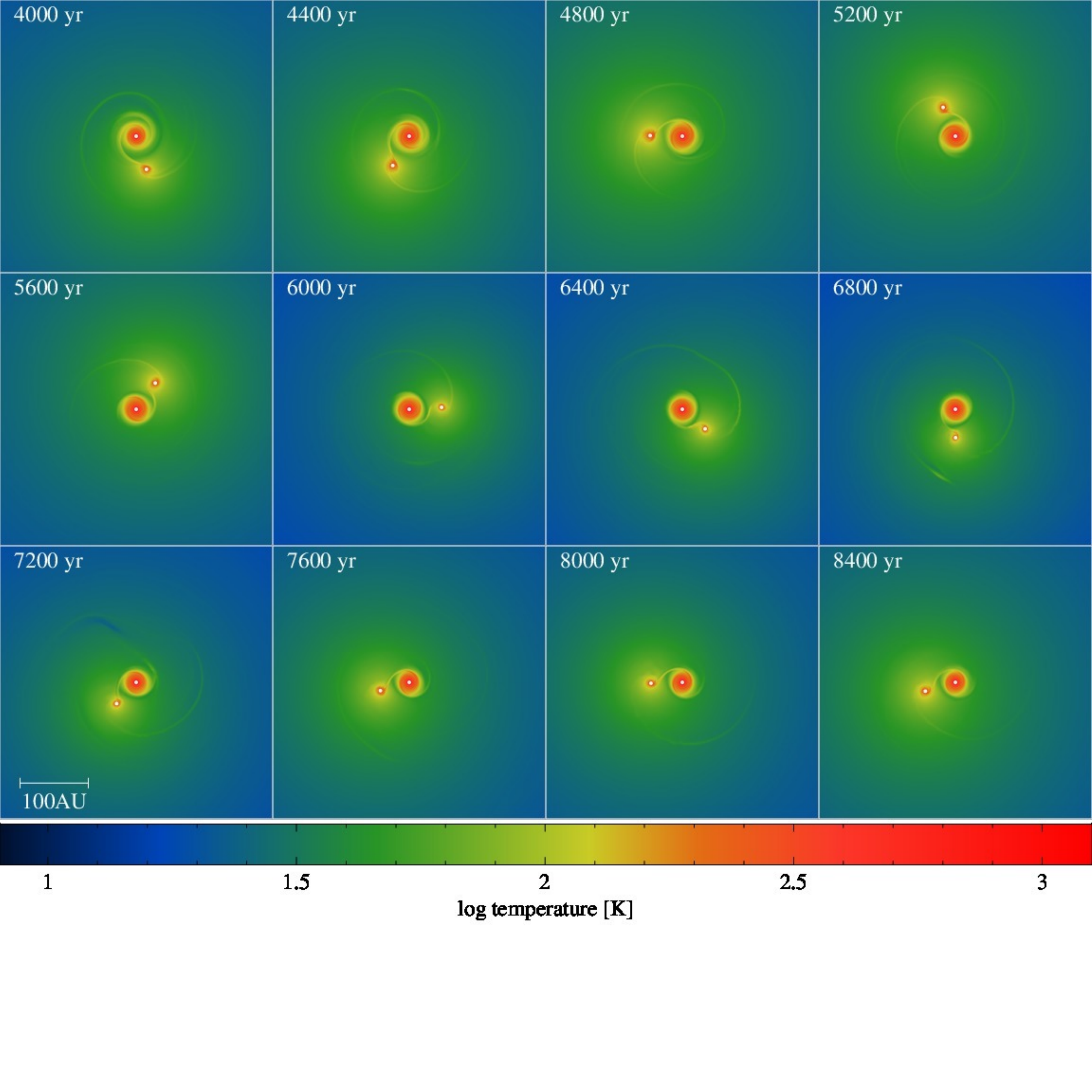}} 
            \caption{Disc evolution with continuous radiative feedback from secondary objects (CRF run). The disc fragments but only one object forms that ends up as a low-mass star.  Radiative feedback from this object suppresses further fragmentation. The object forms on a wide orbit (68~AU) and migrates inwards only by 9 AU within 7.3 kyr.}
            \label{fig:ca1}
            \end{centering}
        \end{figure*}

\begin{figure*}
            \begin{centering}
            \subfloat{\includegraphics[width = 0.75\textwidth, trim = 0cm 6cm 0cm 0cm, clip=true]{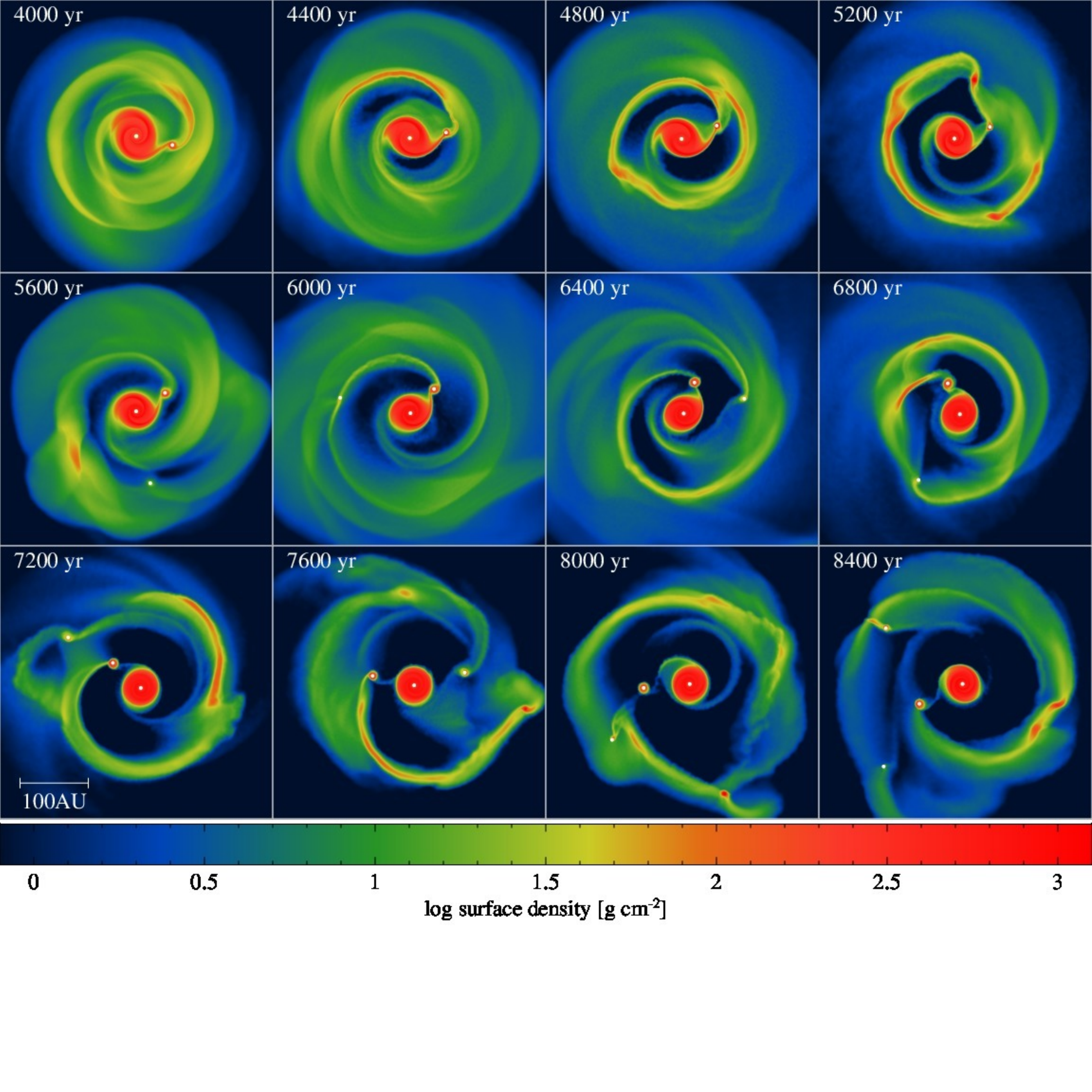}} \\
            \subfloat{\includegraphics[width = 0.75\textwidth, trim = 0cm 6cm 0cm 0cm, clip=true]{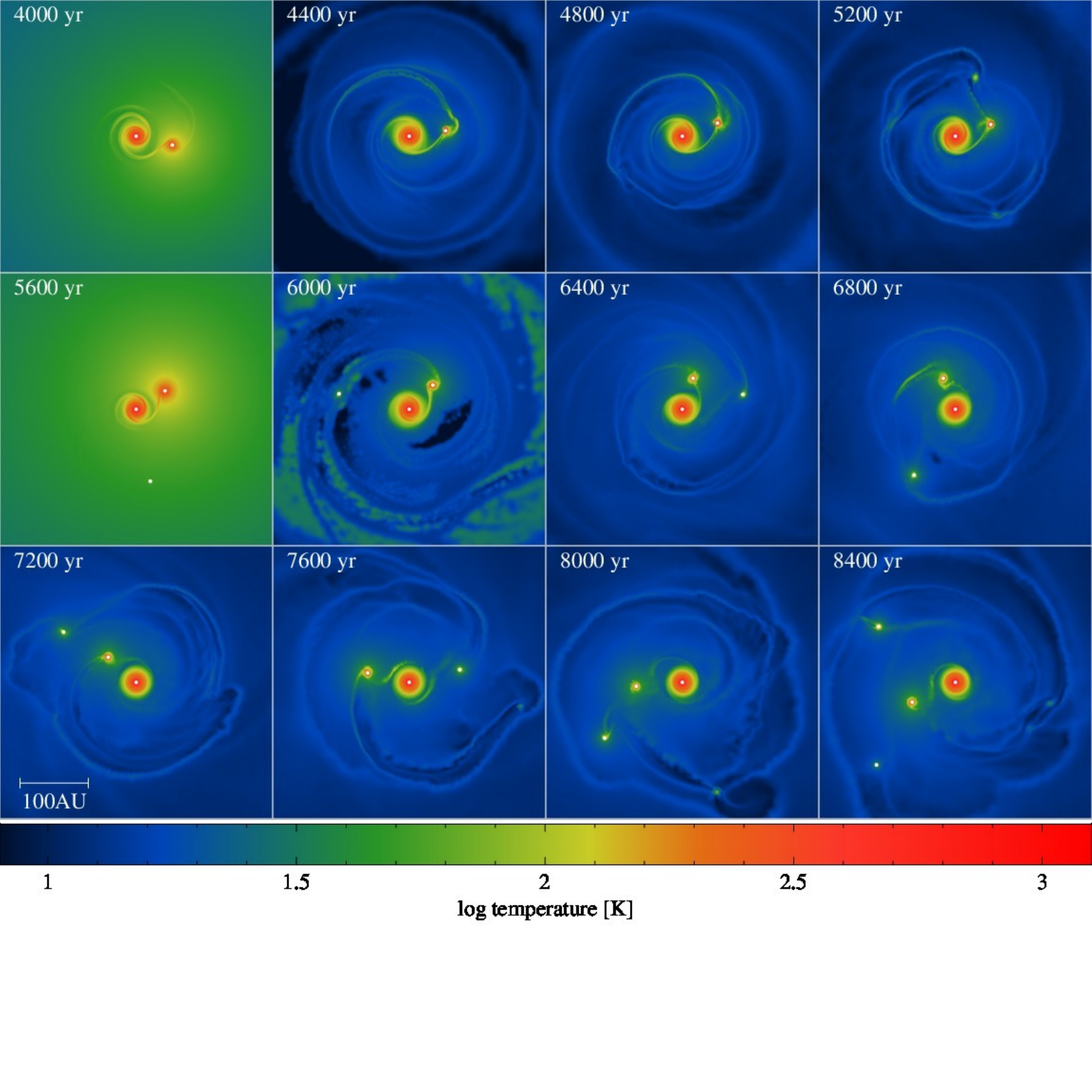}} 
            \caption{Disc evolution  with episodic radiative feedback from secondary objects and a viscosity parameter $\alpha_{\textsc{mri}} = 0.01$ (ERF001 run). The disc fragments and 4 objects form as the disc is cool enough to be gravitationally unstable between accretion episodes.}
            \label{fig:ea1_001}
            \end{centering}
        \end{figure*}

\begin{figure*}
            \begin{centering}
            \subfloat{\includegraphics[width = 0.75\textwidth, trim = 0cm 6cm 0cm 0cm, clip=true]{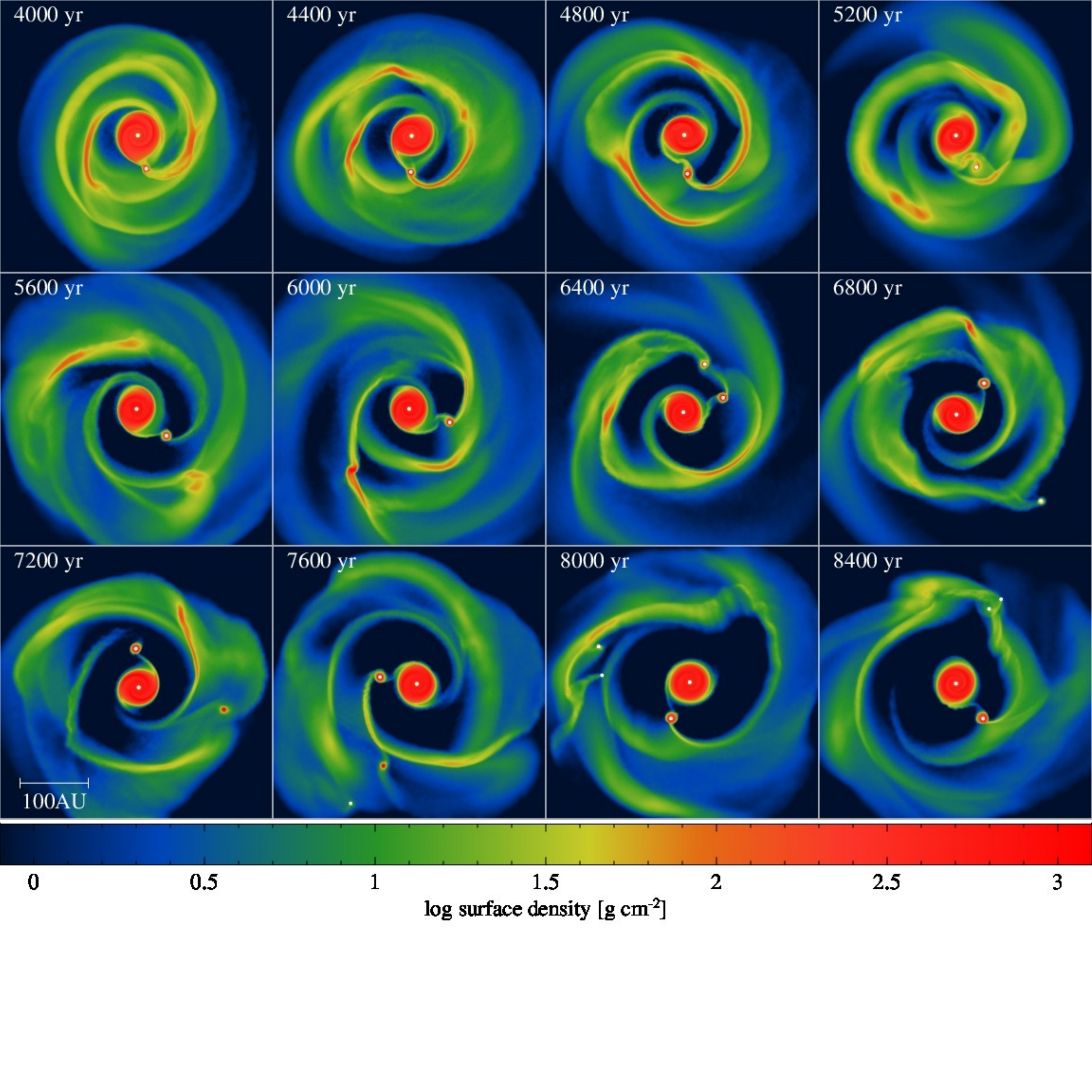}} \\
            \subfloat{\includegraphics[width = 0.75\textwidth, trim = 0cm 6cm 0cm 0cm, clip=true]{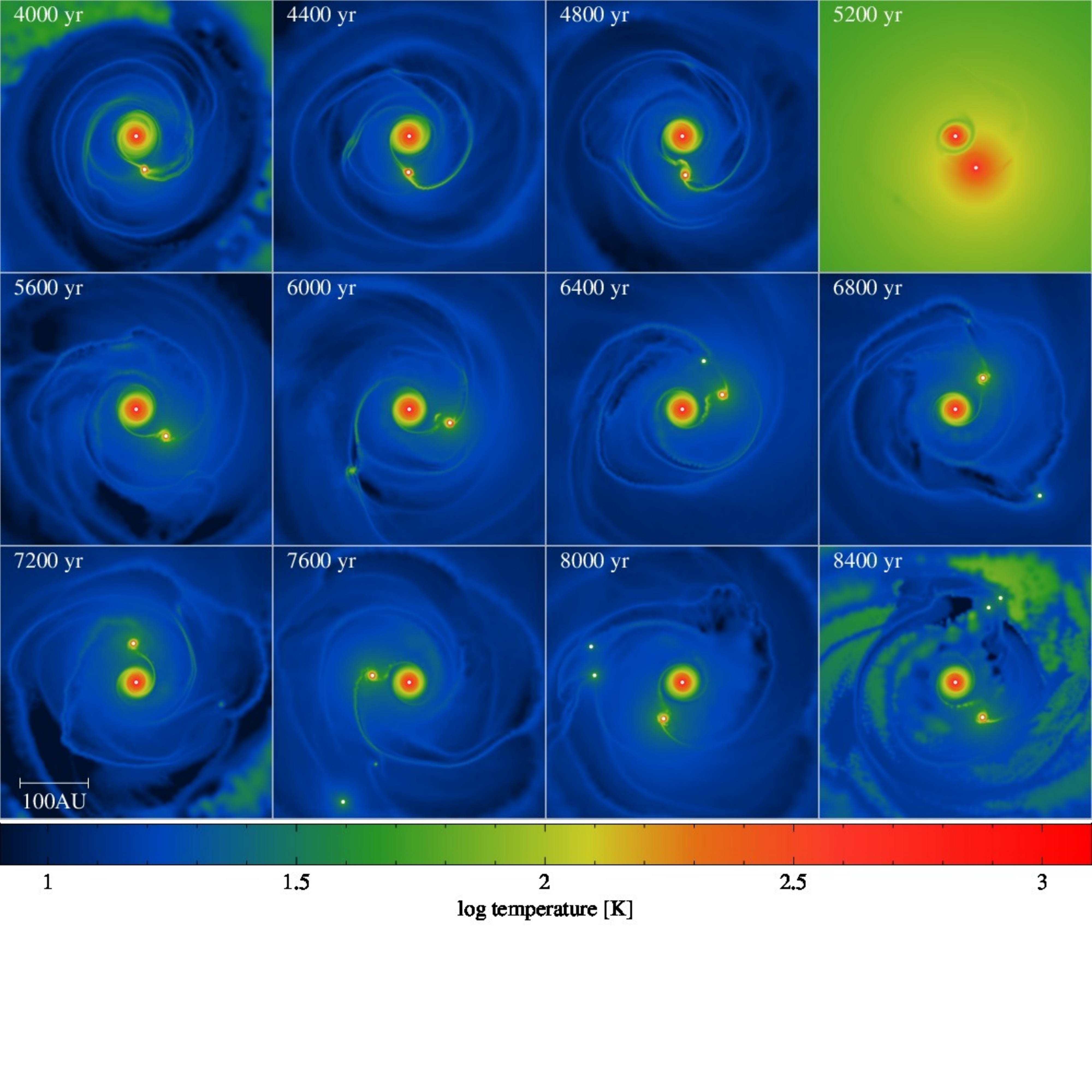}}    
            \caption{Disc evolution  with episodic radiative feedback from secondary objects and a viscosity parameter $\alpha_{\textsc{mri}} = 0.1$ (ERF01 run). The disc fragments and 3 objects form. Two of these objects are planets, as in the ERF001 run.}
            \label{fig:ea1_01}
            \end{centering}
        \end{figure*}

\begin{figure*}
            \begin{centering}
            \subfloat{\includegraphics[width = 0.75\textwidth, trim = 0cm 6cm 0cm 0cm, clip=true]{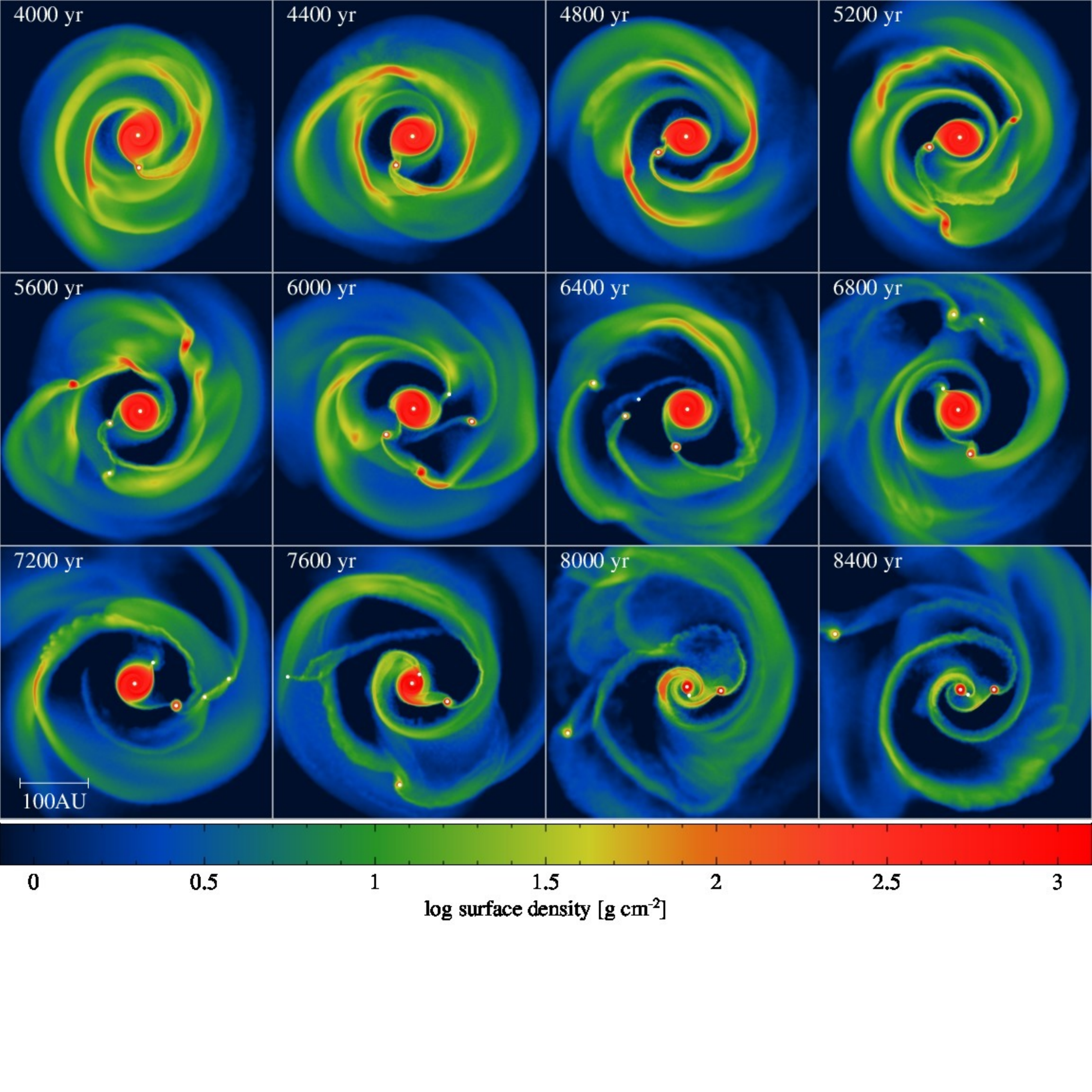}} \\
            \subfloat{\includegraphics[width = 0.75\textwidth, trim = 0cm 6cm 0cm 0cm, clip=true]{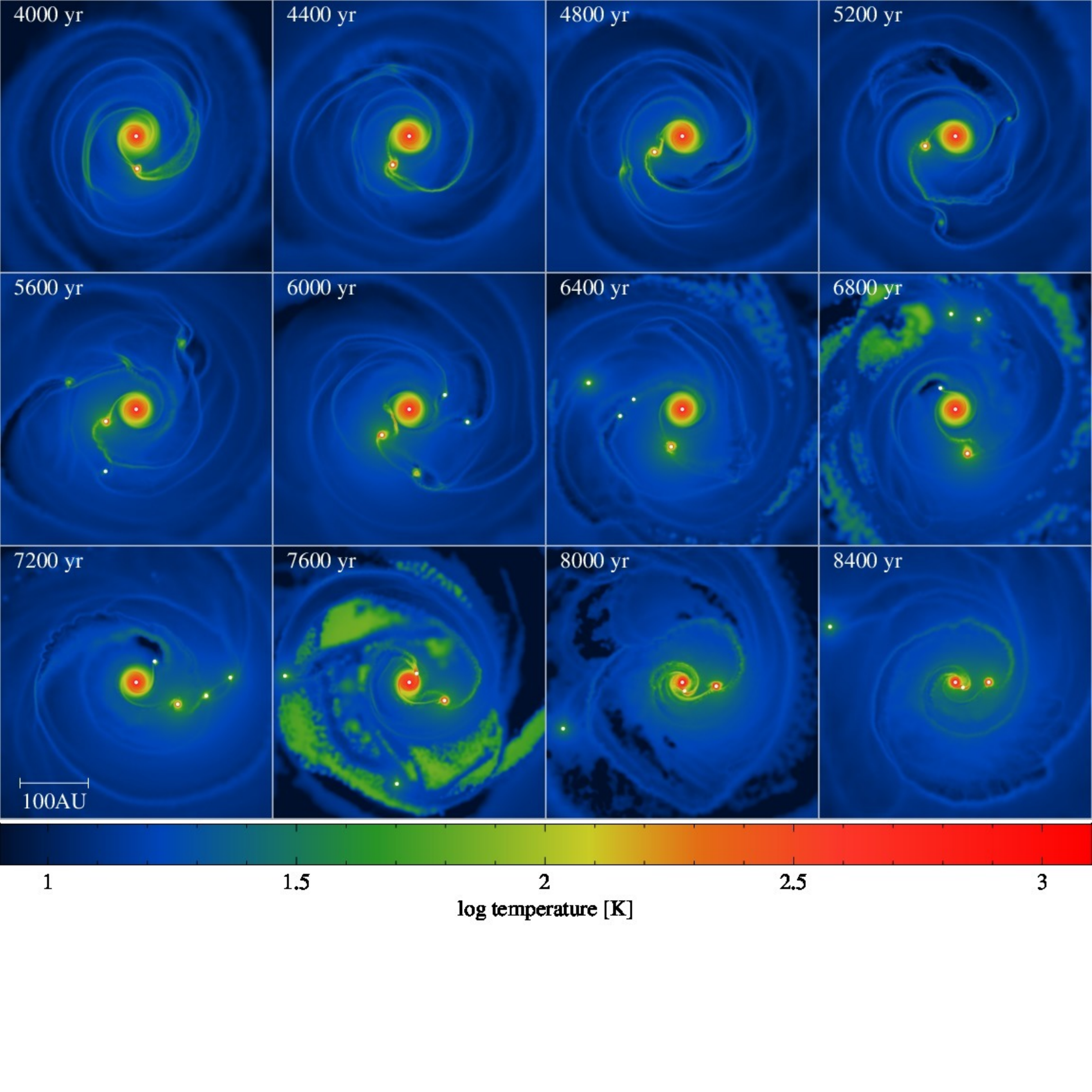}} 
            \caption{Disc evolution  with episodic radiative feedback from secondary objects and a viscosity parameter $\alpha_{\textsc{mri}} = 0.3$ (ERF03 run). The disc fragments and 4 objects form.  One object migrates inwards significantly such that it accretes a large amount of gas while in a close orbit to the central star. The two lowest mass objects are ultimately ejected from the system.}
            \label{fig:ea1_03}
            \end{centering}
        \end{figure*}

    We perform a set of 5 simulations of a $0.3$-M$_{\sun}$  gravitationally unstable protostellar disc around a $0.7$-M$_{\sun}$ star. The disc initial conditions are identical (as described in Section~\ref{sec:initial_conditions}).  The disc is  gravitationally unstable, thus spiral arms form and the disc fragments to form secondary objects in all 5 cases (see Table~\ref{tab:formation}). The only difference between the 5 simulations is the way the radiative feedback from these secondary objects is treated: (i)  for simulation "NRF" there is no radiative feedback from secondary objects, (ii) for simulation "CRF", the radiative feedback from secondary objects is continuous, (iii) for simulations "ERF001", "ERF01", "ERF03" the radiative feedback is episodic. The difference  between the last three simulations is the value of the viscosity parameter due to the MRI, which determines the intensity  and the duration of the outburst (ERF001: $\alpha_{\textsc{mri}} = 0.01$; ERF01: $\alpha_{\textsc{mri}} = 0.1$; ERF03  $\alpha_{\textsc{mri}} = 0.3$). The disc surface density and the disc midplane temperature of the 5 runs are shown in Figures~\ref{fig:na1}-\ref{fig:ea1_03}. In all five simulations the discs evolve identically and at 2.7 kyr an object forms due to gravitational fragmentation. From this point on, the simulations differ as this object provides different radiative feedback in each run. In the NRF run, 7 secondary objects from by disc fragmentation, whereas in the CRF run only 1 secondary object forms. In the ERF runs, 3-4 secondary objects form, i.e. somewhere in between the two previous cases, similarly to what previous studies have found \citep{Stamatellos:2011a,Stamatellos:2012a,Lomax:2014a,Lomax:2015a}. The properties of the objects formed in each run are listed in Table~\ref{tab:formation}. In the next subsections we discuss each of the simulations in detail.

	\begin{table*}
        \centering
        \caption{The properties of objects formed by gravitational fragmentation in the simulations with no radiative feedback from secondary objects (NRF), with continuous radiative feedback (CRF), and with episodic radiative feedback (ERF001, ERF01, ERF03). $N_{\textup{o}}$ is the total number of secondary objects formed, $t_i$ is the formation time of an object, $M_i$ its initial mass, and $M_f$ its final masses (i.e. at the end of the hydrodynamical simulation; $t=10$~kyr). $M_{\textsc{max}}$ is the maximum possible mass it can attain by accreting mass from the disc (see discussion in the text), $\langle \dot{M} \rangle$ is the mean accretion rate, $\dot{M}_{f} $ is the accretion rate onto the object at 10~kyr, $R_i$ is the distance from the star when it forms, $R_f$ is its final distance from the star, and $\Delta R=R_f-R_i$ is its radial migration within 10 kyr. S denotes the central star, LMS  secondary low-mass stars, BD brown dwarfs and P planets. In the final column we mark the boundedness at the end of the NBODY simulation (200~kyr). B and E denote bound and ejected respectively.}
        \resizebox{\textwidth}{!}{\begin{tabular}{l c c c c c c c c c c c c c c}
        	\hline
            \hline
             Run ID & $\alpha_{\textsc{mri}}$ & $\textup{N}_{\textup{o}}$ & $t_{i}$ & $M_{i}$ & $M_{f}$ & $M_{\textsc{max}}$ & $\langle \dot{M} \rangle$ & $\dot{M}_{f}$ & $R_{i}$ & $R_{f}$ & $\Delta R$ & Type & Bound \\

              & & & (kyr) & $\left(\textup{M}_{\textsc{j}}\right)$ & $\left(\textup{M}_{\textsc{j}}\right)$ & $\left(\textup{M}_{\textsc{j}}\right)$ & $\left(10^{-7} \msun \textup{yr}^{-1}\right)$ & $\left(10^{-7} \msun \textup{yr}^{-1}\right)$ & $\left(\textup{AU} \right)$ & $\left(\textup{AU} \right)$ & $\left(\textup{AU} \right)$ & \\
            \hline
            \hline
            NRF & - & 7 & \scell{0.0 \\ 2.7 \\4.3 \\ 5.5 \\ 5.9 \\ 6.0 \\ 7.1 \\ 7.5} &
                                  \scell{733 \\ 2 \\ 4 \\ 2 \\ 2 \\ 2 \\ 1 \\ 1} &
                                  \scell{773 \\ 97 \\ 48 \\ 13 \\ 4 \\ 66 \\ 4 \\ 3} &
                                  \scell{774 \\ 99 \\ 53 \\ 13 \\ 4 \\ 67 \\ 4 \\ 3} &
                                  \scell{37.7 \\ 124 \\ 75.0 \\ 22.0 \\ 4.63 \\ 154 \\ 7.78 \\ 6.33} &
                                  \scell{5.29 \\ 14.7 \\ 27.3 \\ 2.83 \\ 0.02 \\ 4.57 \\ 3.19 \\ 0.34} &
                                  \scell{0 \\ 77 \\ 65 \\ 160 \\ 270 \\ 103 \\ 191 \\ 103} &
                                  \scell{0 \\ 105 \\ 25 \\ 144 \\ 570 \\ 15 \\ 169 \\ 235} & 
                                  \scell{0 \\ 28 \\ -40 \\ -16 \\ 300 \\ -88 \\ -22 \\ 132} & 
                                  \scell{S \\ LMS \\ BD \\ P/BD \\ P \\ BD \\ P \\ P} &
                                  \scell{- \\ B \\ E \\ E \\ E \\ B \\ E \\ E} \\ \\

            CRF & - & 1 &         \scell{0.0 \\ 2.7} & 
                                  \scell{733 \\ 2} & 
                                  \scell{772 \\ 79} & 
                                  \scell{826 \\ 191} &
                                  \scell{36.8 \\ 100} &
                                  \scell{31.9 \\ 66.6} &
                                  \scell{0 \\ 77} & 
                                  \scell{0 \\ 68} &
                                  \scell{0 \\ -9} &
                                  \scell{S \\ BD/LMS} &
                                  \scell{- \\ B} \\ \\

            ERF001   & 0.01 & 4 & \scell{0.0 \\ 2.7 \\ 5.4 \\ 8.2 \\ 9.8} & 
                                  \scell{733 \\ 2 \\ 3 \\ 2 \\ 2} &
                                  \scell{772 \\ 87 \\ 32 \\ 8 \\ 3} &
                                  \scell{811 \\ 127 \\ 42 \\ 32 \\ 5} &
                                  \scell{37.2 \\ 111 \\ 60.5 \\ 31.0 \\ 51.0} &
                                  \scell{33.0 \\ 35.0 \\ 8.54 \\ 20.7 \\ 2.01} &
                                  \scell{0 \\ 77 \\ 93 \\ 166 \\ 119} &
                                  \scell{0 \\76 \\ 178 \\ 97 \\ 104} &
                                  \scell{0 \\ -1 \\ 85 \\ -69 \\ -15} &
                                  \scell{S \\ LMS \\ BD \\ P \\ P} &
                                  \scell{- \\ B \\ B \\ E \\ E} \\ \\

            ERF01   & 0.1  & 3 &  \scell{0.0 \\ 2.7 \\ 6.3 \\ 7.9} &
                                  \scell{733 \\ 2 \\ 3 \\ 2} &
                                  \scell{771 \\ 91 \\ 13 \\ 9} &
                                  \scell{805 \\ 124 \\ 63 \\ 25} &
                                  \scell{37.3 \\ 117 \\ 26.9 \\ 31.4} &
                                  \scell{25.2 \\ 24.3 \\ 36.4 \\ 11.9} &
                                  \scell{0 \\ 77 \\ 85 \\ 137} &
                                  \scell{0 \\ 64 \\ 123 \\ 129} & 
                                  \scell{0 \\ -13 \\ 38 \\ -8} &
                                  \scell{S \\ LMS \\ P/BD \\ P} &
                                  \scell{- \\ B \\ E \\ E} \\ \\

            ERF03   & 0.3  & 4 &  \scell{0.0    \\ 2.7  \\ 5.5  \\ 6.0  \\ 6.0} &
                                  \scell{733    \\ 2    \\ 2    \\ 2    \\ 3} &
                                  \scell{771    \\ 105  \\ 66   \\ 17   \\ 16} &
                                  \scell{782    \\ 116  \\ 73   \\ 24   \\ 16} &
                                  \scell{44.3   \\ 135  \\ 138  \\ 34.9 \\ 72.0} &
                                  \scell{33.6   \\ 31.2 \\ 18.5 \\ 29.5 \\ 0.0} &
                                  \scell{0      \\ 77   \\ 106  \\ 87   \\ 96} &
                                  \scell{0      \\ 40   \\ 14   \\ 142  \\ 300} &
                                  \scell{0      \\ -37  \\ -92  \\ 55   \\ 204} & 
                                  \scell{S      \\ LMS  \\ BD   \\ BD   \\ BD} &
                                  \scell{-      \\ B    \\ B    \\ E    \\ E} \\ \\
            \hline
            \end{tabular}}
            \label{tab:formation}
        \end{table*}

    \subsection{No radiative feedback (NRF)} 
    \label{sub:no_radiative_feedback}

\begin{figure*}
            \begin{centering}
            \subfloat{\includegraphics[width = 0.75\textwidth, trim = 0cm 6cm 0cm 0cm, clip=true]{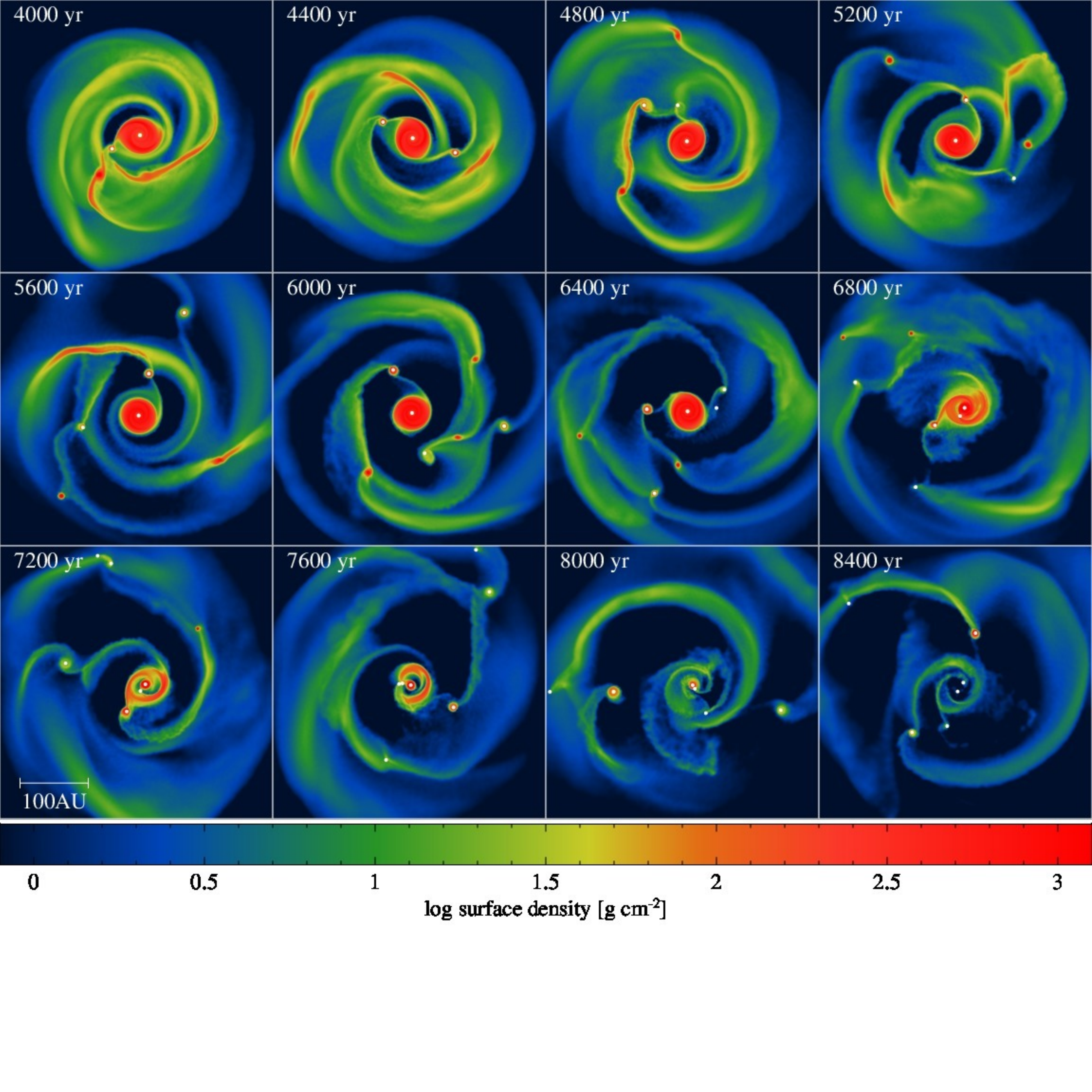}} \\
            \subfloat{\includegraphics[width = 0.75\textwidth, trim = 0cm 6cm 0cm 0cm, clip=true]{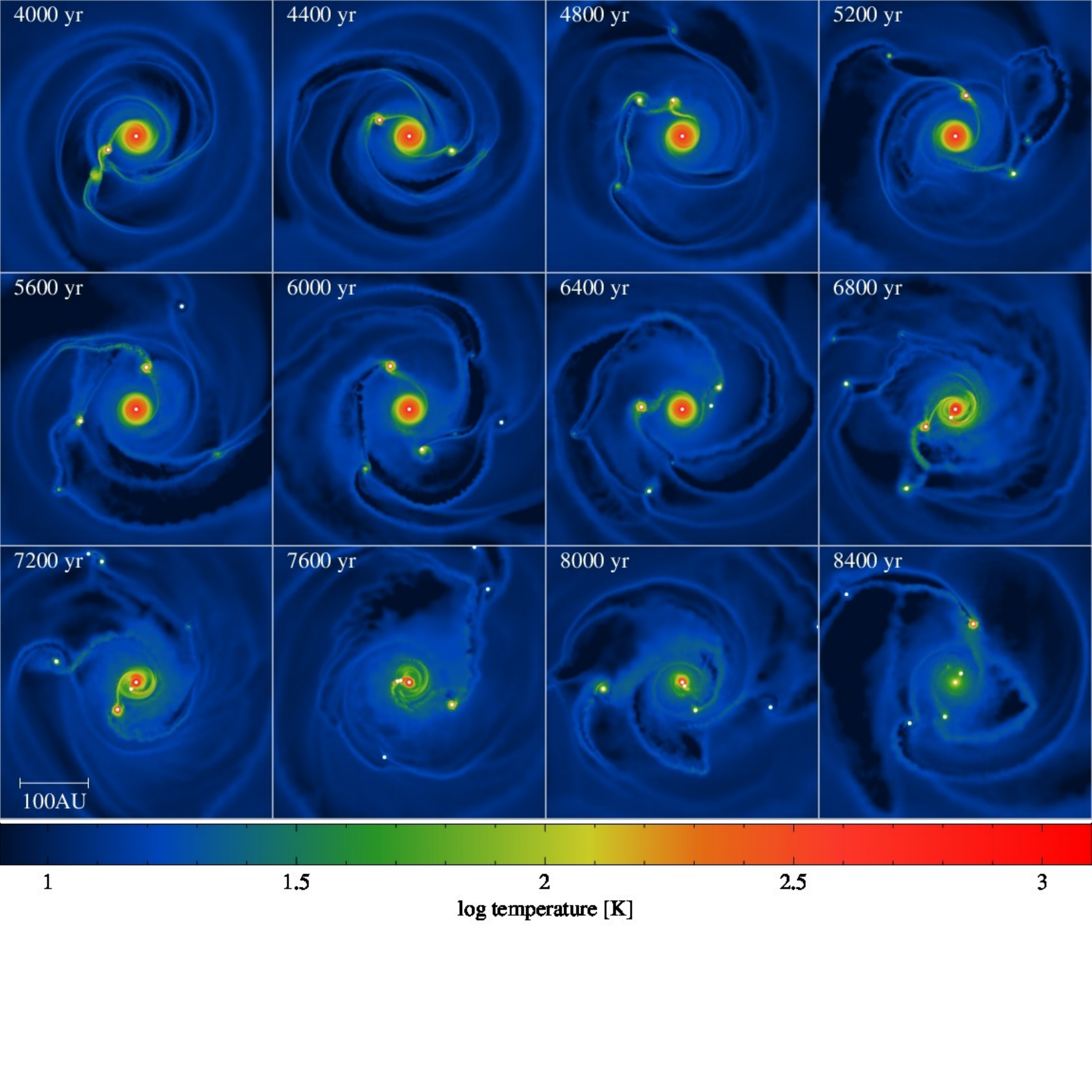}} 
            \caption{Disc evolution  without any radiative feedback from secondary objects (NRF run). The top snapshots show the disc surface density  and the bottom snapshots show the disc midplane temperature (at times as marked on each  graph). 7 objects form by gravitational fragmentation due to the disc cooling fast enough in its outer regions. Most of the objects formed are brown dwarfs and planets. Planets are ultimately ejected from the system.}
            \label{fig:na1}
            \end{centering}
        \end{figure*}

        Figure \ref{fig:na1} shows the evolution of the surface density and midplane temperature for the disc whereby no radiative feedback is provided from secondary objects that form in the disc. Spiral arms develop and the disc fragments to form 7 secondary objects (see Table~\ref{tab:formation}).  Fragmentation occurs outside $65$~AU where the disc is gravitationally unstable and cools fast enough \citep[e.g.][]{Rice:2003e,Rice:2005a,Stamatellos:2007c,Stamatellos:2011d}. After 10 kyr, the first of these objects has accreted a sufficient amount of gas to become a low-mass hydrogen-burning star ($M=97$~M$_{\rm J}$). Two brown dwarfs are formed (with masses 48 and 66~M$_{\rm J}$)  and orbit within 25~AU of the central star (at 25 and  15 AU, respectively). Three of the objects formed  remain in the planetary-mass regime. These form at a late stage and at large orbital radii, thus having less time to accrete gas from the disc. One of these planets undergoes a net radial outward migration of 300~AU between its formation at 5.9 kyr and the end of the hydrodynamical simulation (10~kyr). These objects are bound to the central star by the end of the hydrodynamic simulation. However, a few of them are loosely bound at large radii ($R>150$~AU for 3 of them), and therefore destined to be ejected from the system. Indeed, at the end of the NBODY calculation (at 200~kyr), all but two of these objects are ejected from the system, becoming free-floating planets and brown dwarfs \citep[see also][]{Stamatellos:2009a,Li:2015b, Li:2016a, Vorobyov:2016a}.
        
    \subsection{Continuous radiative feedback (CRF)} 
    \label{sub:continuous_radiative_feedback}

        Figure \ref{fig:ca1} shows the evolution of the disc surface density and disc midplane temperature for the simulation with continuous radiative feedback from secondary objects that form in the disc. The disc fragments but only one secondary object forms. Continuous radiative feedback from this object heats and stabilises the disc; therefore, no further fragmentation occurs. The object carves out a gap within the disc and migrates inwards  9 AU by the end of the hydrodynamical simulation (i.e. within 7.3 kyr since its formation). At this point it has accreted enough gas to become a high-mass brown dwarf and is close to overcoming the hydrogen-burning mass limit. As there is still  plenty of gas within the disc, the ultimate fate of this system is a binary comprising a solar-type and a low-mass secondary star.
    

    \subsection{Episodic radiative feedback (ERF)} 
    \label{sub:episodic_radiative_feedback}
 
        Figures \ref{fig:ea1_001}, \ref{fig:ea1_01} and \ref{fig:ea1_03} show the surface density and disc midplane temperature evolution for the simulations with episodic radiative feedback from secondary objects forming in the disc, in three different cases: $\alpha_{\textsc{mri}} = 0.01, 0.1, 0.3$, respectively. The disc fragments as in the previous cases; the  radiative feedback from secondary objects is now episodic due to episodic accretion. During the accretion/outburst episodes, the mass that has accumulated in the inner disc region of a secondary object flows onto the object, resulting in an increase of its accretion luminosity that affects a large portion of the disc around the central star. This is evident by the sudden increase in the temperature (e.g.  in  Figures \ref{fig:ea1_001} and \ref{fig:ea1_01}). The increase of the temperature in the disc is three- to four-fold (see Figure~\ref{fig:QTCD_Comparison}b), which is enough to stabilise the disc during the outburst. However, in all three cases, when the outburst stops the disc becomes unstable and fragments.
        
        The number of secondary objects formed is similar in all three cases (3-4 objects). Therefore, fewer objects form than in the non-radiative feedback case and more objects than the continuous radiative feedback case \citep{Stamatellos:2011a,Stamatellos:2012a,Lomax:2014a,Lomax:2015a}.

        The frequency and duration (see Table~\ref{tab:episodicDuration}) of the accretion/feedback outbursts are important for the  gravitational stability of the disc. The total duration of episodic outbursts drops from $\sim18\%$ to $\sim0.8\%$ of the simulated disc lifetime (10~kyr), as the viscosity parameter $\alpha_{\textsc{mri}}$ is increased from $0.01$ to $0.3$. A larger $\alpha_{\textsc{mri}}$ results in stronger but shorter outbursts. The number of secondary objects forming in the disc does not strongly depend on $\alpha_{\textsc{mri}}$, which indicates that for suppressing disc fragmentation the total duration of episodic outbursts must be longer. 
        
        We find that the average mass of secondary objects at  the end of the hydrodynamical simulation (10~kyr) increases with $\alpha_{\textsc{mri}}$; the average masses are 33, 38 \& 51~$\textup{M}_{\textsc{j}}$ for $\alpha_{\textsc{mri}} =$ 0.01, 0.1,  and 0.3, respectively.  In all cases, the two lowest mass objects are ultimately ejected from the system. For $\alpha_{\textsc{mri}} = 0.01$, the two lowest mass objects are planets. For $\alpha_{\textsc{mri}} = 0.1$, the two lowest mass objects consist of a planet and a brown dwarf. And finally, for $\alpha_{\textsc{mri}} = 0.3$, the two lowest mass objects are brown dwarfs. The estimated maximum mass that all  of all these objects will eventually attain (see next section) is above the deuterium-burning limit, except for one object in the $\alpha_{\textsc{mri}} = 0.01$ run.

        We further find that subsequent formation of secondary objects happens on a more rapid timescale for greater values of $\alpha_{\textsc{mri}}$. Radiative feedback episodes are shorter for a higher $\alpha_{\textsc{mri}}$: the disc cools fast after an episode ends, becoming gravitationally unstable and fragments again within a shorter time.

        \begin{center}
            \begin{table}
            \centering
                \caption{The duration of episodic accretion events from each secondary object in the simulations which consider episodic radiative feedback.}
                \begin{tabular}{l c c c c c}
                    \hline
                    \hline

                   Run ID & $\alpha_{\textsc{mri}}$ & Sink \# & Episodes & Duration (yr) \\

                    \hline
                    \hline

                    ERF001 & 0.01 & 2   & 3 & 1170 \\
                           &      & 3   & 2 & 487  \\
                           &      & 4   & 1 & 90   \\
                           &      & 5   & 0 & 0    \\
                           &      & All & 6 & 1747 \\ \\

                    ERF01  & 0.1  & 2   & 3 & 108  \\
                           &      & 3   & 1 & 10   \\
                           &      & 4   & 1 & 9    \\
                           &      & All & 5 & 128  \\ \\

                    ERF03  & 0.3  & 2   & 3 & 46 \\
                           &      & 3   & 2 & 22  \\
                           &      & 4   & 1 & 4  \\
                           &      & 5   & 2 & 9  \\
                           &      & All & 6 & 81 \\

                    \hline
                \end{tabular}
                \label{tab:episodicDuration}
            \end{table}
        \end{center}
    

    \subsection{Comparison of simulations with different radiative feedback from secondary objects} 
    \label{sub:comparison_of_simulations}

        Table \ref{tab:formation} lists information pertaining to the secondary objects that form in the disc simulations. We list the number of objects formed in each simulation, their initial and final masses (i.e. at the end of the hydrodynamical simulation, $t=10$~kyr), and an estimate of the maximum mass they can ultimately attain (considering that they will still be evolving in a gaseous disc), the gas accretion rate onto them, their formation and final radius, their type, and their estimated boundedness at the end of the NBODY simulation (200~kyr).
        
        The maximum mass, $M^i_{\textsc{max}}$, that an object $i$ can attain is calculated as follows. We assume that each object will continue to accrete at its accretion rate $\left(\dot{M}^i_{f}\right)$ at the end of the hydrodynamic simulation (which is likely an overestimate as generally the accretion rate decreases, unless there is some kind of violent interaction within the disc). Therefore, the maximum mass that an object $i$ can attain is:
        \begin{equation}
            \label{eqn:theoreticalMaximumMass}
        M^i_\textsc{max}=M^i_f+\dot{M}^i_{f}\ t_{acc}\,,
        \end{equation}
        where $M^i_{f}$ is the mass of an object $i$ at $t=10$~kyr, and $t_{acc}$ is the time for which it will keep on accreting gas. We also assume that only a fraction $\xi=0.9$ of all the gas from the disc could eventually accrete either onto the central star or onto the secondary objects, therefore
        \begin{equation}
            \xi M_{\textup{disc}}= \sum_{sec}{M^i_{f}} + \sum_{all}{\dot{M}_f^{i}t_{acc}}
        \label{eqn:finalMassEquate}
        \end{equation}
        where the first sum on the right hand side is over the secondary objects and the second sum is over all objects to include gas accreting onto the central star. We assume that the accretion time $t_{acc}$ is the same for all objects within each simulation, therefore it is calculated such that    
        \begin{equation}
            t_{acc} = \frac{\xi M_{\textup{disc}} - \sum_{sec}M_f^{i}}{\sum_{all}\dot{M}_f^{i}}.
        \label{eqn:accretionTime}
        \end{equation}
        The maximum estimated mass for an object $i$ can then be calculated using Equation~\ref{eqn:theoreticalMaximumMass}.


    \begin{figure}
        \begin{center}
            \includegraphics[width = 0.5\textwidth, trim = 0cm 0cm 0cm 0cm, clip=true]{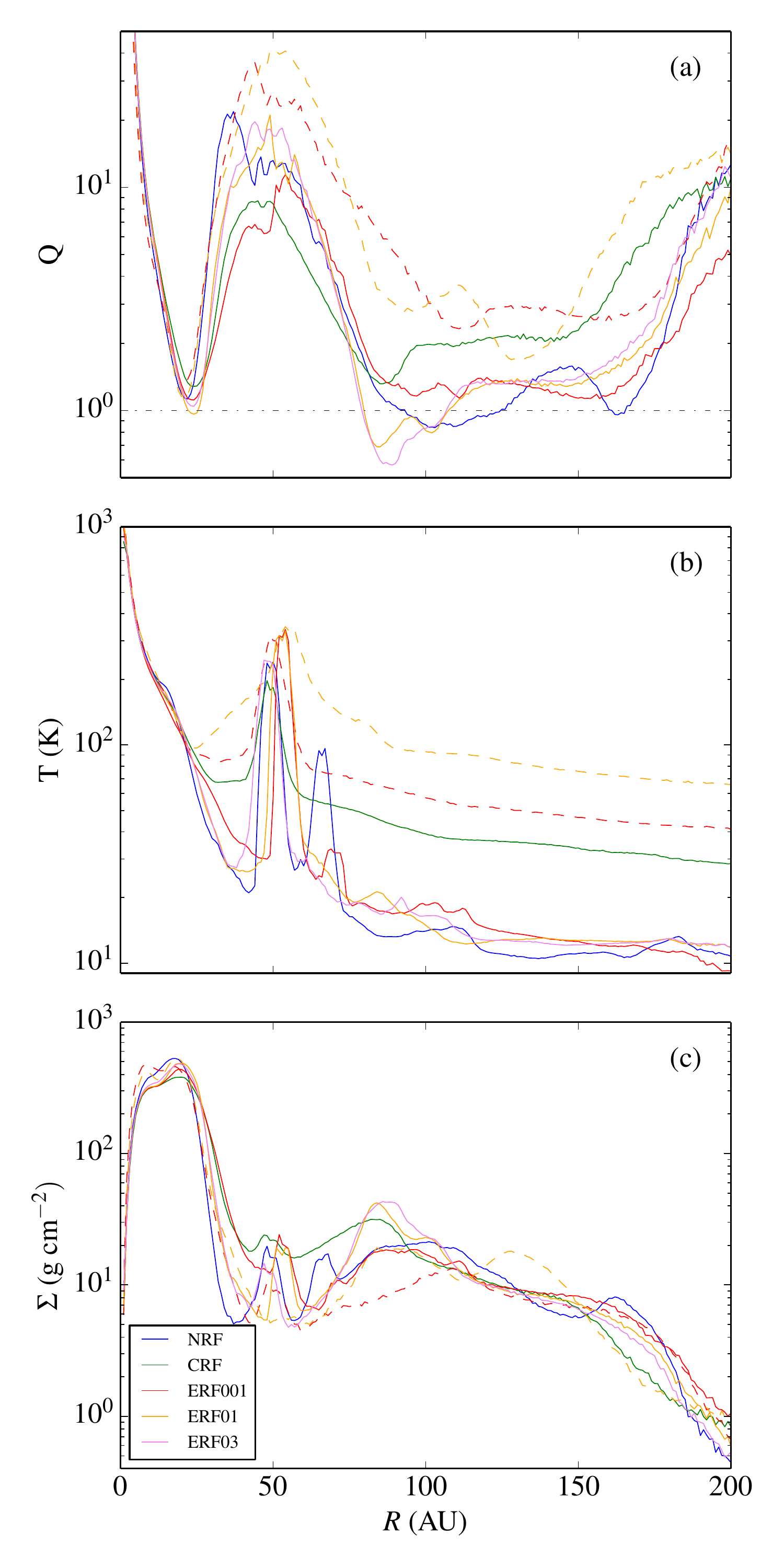}
            \caption{(a) Azimuthally averaged Toomre parameter $Q$, (b) disc midplane temperature, and (c) disc surface density for all simulations at  $t=4.4$~kyr. The coloured dashed lines correspond to times when outburst episodes are happening: $t=5.6$~kyr and $t=5.2$~kyr for the simulations ERF001 and ERF01, respectively. The disc inner region is gravitationally stable due to the high temperature, whereas the disc is unstable outside $\sim 70$~AU. The temperature peaks between 50 and 100~AU correspond to regions close to secondary objects.}
            \label{fig:QTCD_Comparison}
        \end{center}
    \end{figure}

    The number of secondary objects that form in the disc is strongly affected by the type of their radiative feedback. We find that without radiative feedback, 7 objects form within the disc (Figure \ref{fig:na1}) compared to only one object forming where continuous radiative feedback is considered (Figure \ref{fig:ca1}). Continuous radiative feedback from the secondary object raises the disc temperature enough  (see Figure~\ref{fig:QTCD_Comparison}b) to stabilise the disc  (Figure~\ref{fig:QTCD_Comparison}a) and suppress further fragmentation. The number of objects when episodic radiative feedback is considered is in between the two previous cases ($3-4$ objects). This behaviour has been seen in previous simulations \citep{Stamatellos:2011a,Stamatellos:2012a,Lomax:2014a,Lomax:2015a}. 

\begin{figure}
        \begin{center}
            \includegraphics[width = \columnwidth, trim = 0cm 0cm 0cm 0cm, clip=true]{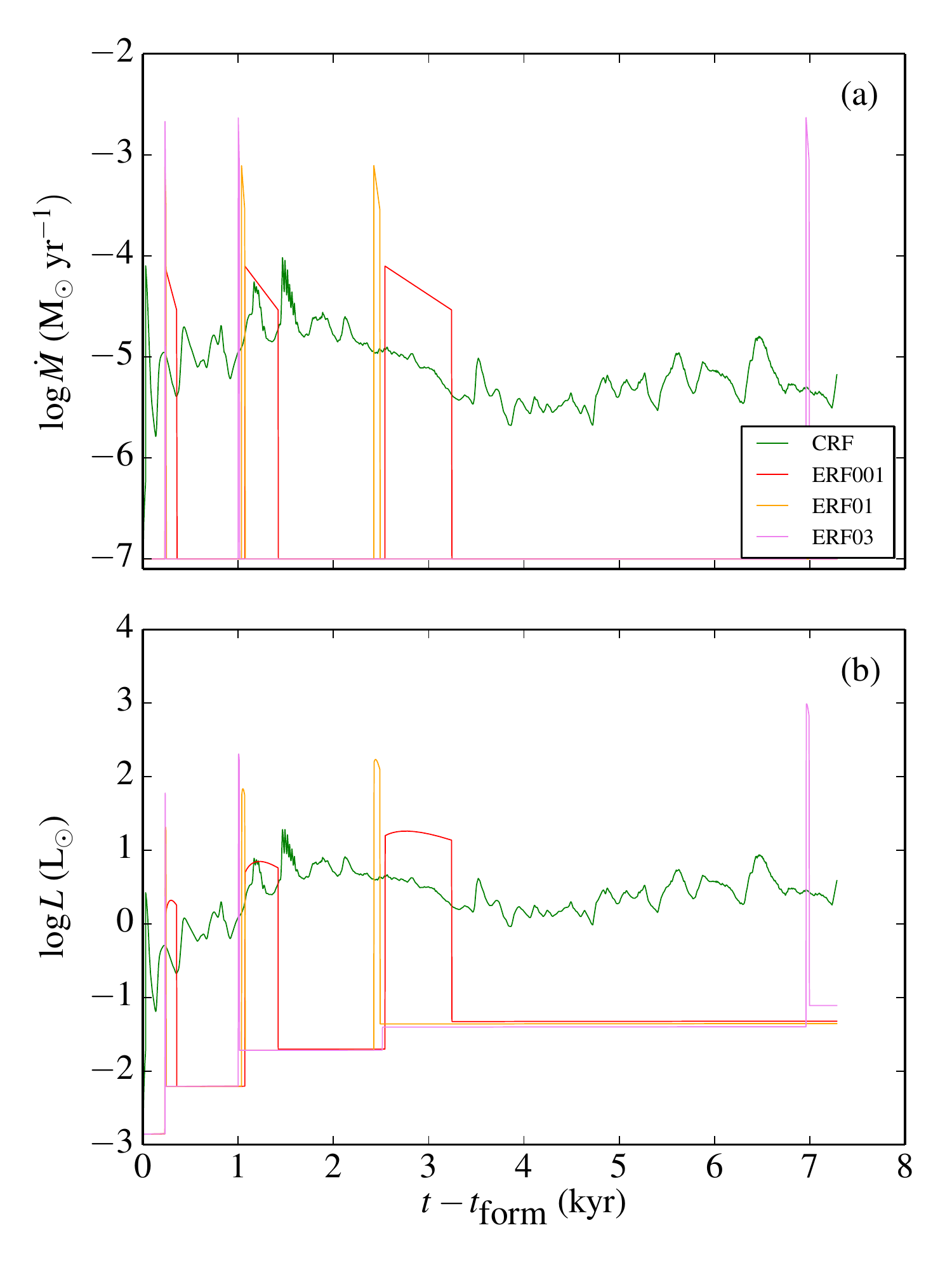}
            \caption{ (a) Mass accretion rates onto and (b) accretion luminosities of the first secondary object that forms in each of the simulations where radiative feedback is considered. Time is given with respect to the formation time of each object. At their initial stages of their formation, secondary objects are high accretors and they may even outshine the central star. In the case where radiative feedback is episodic this only happens for a short time and therefore would be difficult to observed. In the case of continuous radiative feedback (which is probably not realistic) secondary objects may outshine the central star for longer.}
            \label{fig:luminosity_accretion}
        \end{center}
    \end{figure}

Radiative feedback from secondary objects affects the entire disc as these secondary objects are high accretors  (at their initial stages of their formation). For a short time they may even outshine the central star (see Figure~\ref{fig:luminosity_accretion}).  The assumed pseudo-background temperature profile provided by each secondary object  (see Equation~\ref{eqn:Tbgr}) influences the temperature at a given location in the disc and may affect disc fragmentation \citep{Stamatellos:2011d} but probably not significantly. If we adopt a pseudo-background temperature profile with $q=3/4$ instead of $q=1/2$, then the disc temperature at a distance 50 AU from a radiative object will be a factor of $\sim5$ smaller, and the Toomre parameter $Q$ (see Figure~\ref{fig:QTCD_Comparison}a) a factor of $\sim2$ smaller, bringing it (for the CRF and ERF runs), close to the critical value for fragmentation  \citep[$Q\approx1$; see e.g.][]{Durisen:2007a}. However, this is the maximum expected effect. Even in the case of  $q=3/4$  (which is an upper limit for $q$) the disc temperature is expected to be higher than the minimum "background" value due to energy dissipation within the disc as it is  gravitationally unstable.

    With regard to the episodic radiative feedback runs, the number of secondary object does not vary much  for a different MRI viscosity parameter $\alpha_{\textsc{mri}}$. 4 objects form when $\alpha_{\textsc{mri}} = 0.01$; 3 objects form when $\alpha_{\textsc{mri}} = 0.1$; 4 objects form when $\alpha_{\textsc{mri}} = 0.3$. More secondary objects result in more radiative feedback episodes and a hotter disc for longer periods of time. Thus this provides sustained stability against gravitational fragmentation. The duration of episodic outbursts affects the stability of the disc. For a smaller $\alpha_{\textsc{mri}}$, episodes are longer and provide longer periods of stability. The opposite is true for a larger $\alpha_{\textsc{mri}}$. This is shown in Table \ref{tab:episodicDuration}. However, it is evident that episodic feedback from only 1 or 2 secondary objects cannot suppress further disc fragmentation, in contrast with the continuous feedback case, where the presence of just one secondary object suppresses fragmentation.

Observations of episodic outbursts from secondary objects do not require high-sensitivity; during these outbursts their luminosity increases from a few $\textup{L}_{\sun}$ to tens of $\textup{L}_{\sun}$ (see Figure~\ref{fig:luminosity_accretion}). In the case of $\alpha_{\textsc{mri}} = 0.01$, where the outburst events are mild and long, 18\% of the initial 10~kyr of the disc's lifetime correspond to the outburst phase. On the other hand, when $\alpha_{\textsc{mri}} = 0.3$, where  the events are short and intense, this percentage drops down to just 0.8\% (see Table~\ref{tab:episodicDuration}). However, episodic accretion events are expected to be relatively more frequent only during the initial stages of disc evolution, i.e. within a few kyr after the disc's formation, while the newly formed secondary objects are vigorously accreting gas from the disc. Therefore, such outbursts from secondary objects  at the initial stages of disc evolution should not significantly influence the observed number of outbursting sources. \cite{Scholz:2013a} observed a sample of $\sim4000$ YSOs over a period of 5 years and they found $1-4$ possible outbursting sources indicating that outbursts happens at intervals of $(5-50)$~kyr; this is roughly consistent with our models after the initial $\sim 4$~kyr during the disc's evolution (see Figure~\ref{fig:luminosity_accretion}).

    Figure \ref{fig:QTCD_Comparison} shows a comparison between radially-averaged Toomre parameter, temperature and surface density for a representative snapshot from each simulation exhibiting strong spiral features ($t=4.4$~kyr). Within the inner $\sim 25$~AU, the disc is stable due to heating from  the central star. The peaks in surface density and temperature around $\sim 50$~AU correspond to regions around secondary objects. The discs are unstable or close to being unstable outside $\sim 80$~AU in all cases apart from the CRF run and the ERF runs (during episodic outbursts).

    \begin{figure}
        \begin{centering}
            \includegraphics[width = 0.5\textwidth, trim = 0cm 0cm 0cm 0cm, clip=true]{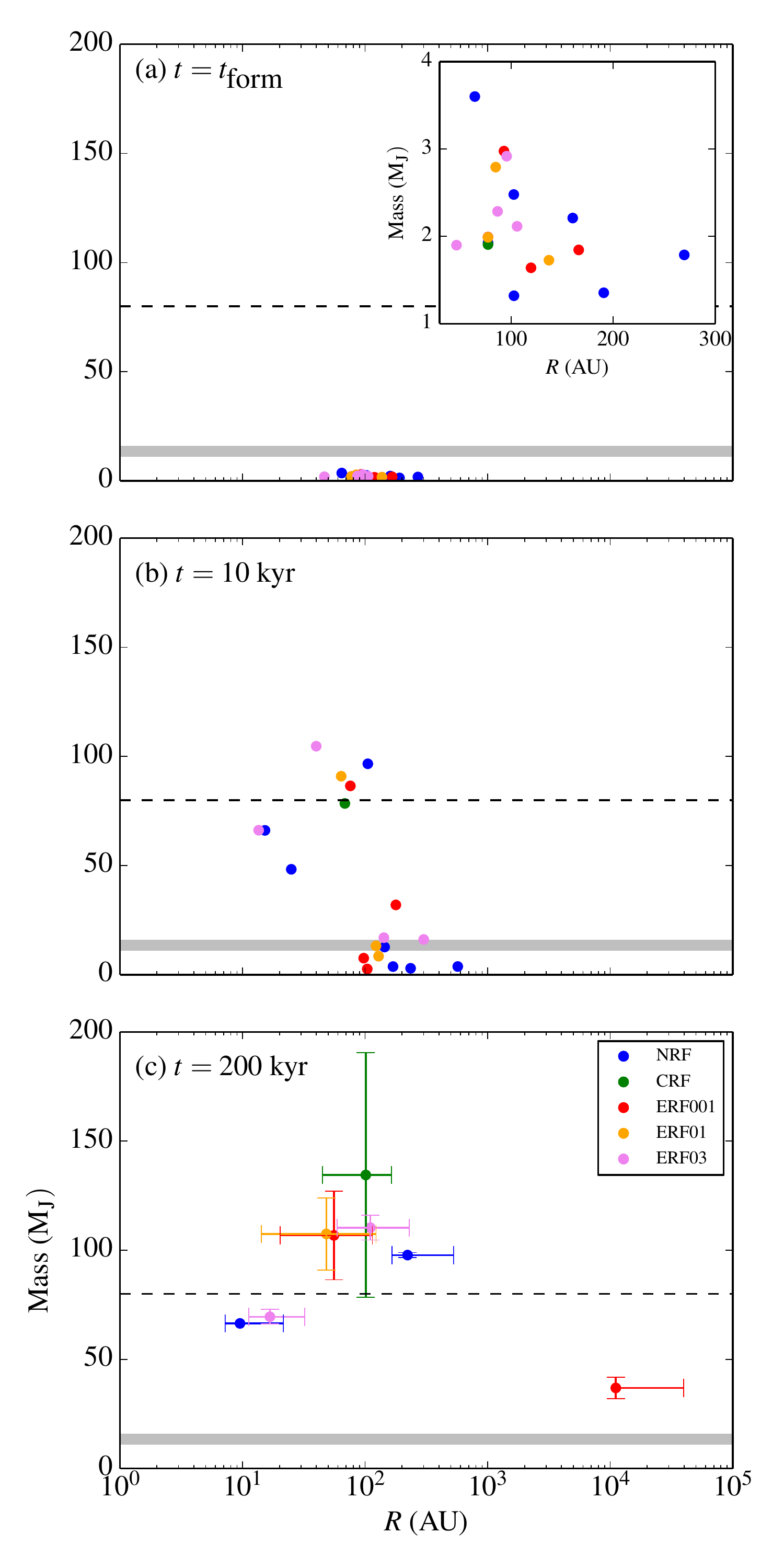}
            \caption{Mass-radius plots of the secondary objects formed by disc fragmentation in all 5 simulations. (a) Mass and radius at formation. A zoomed inset panel is shown for clarity. (b) Mass and radius at the end of the hydrodynamical simulation (10 kyr). (c) Mass and  and semi-major axis at the end of the NBODY simulation (200 kyr). The upper mass limits correspond to the maximum mass that the object may attain (see text for details), whereas the lower mass limits corresponds to the mass of the object at the end of the hydrodynamical simulation. The horizontal bars in this panel represent the periastron and apoastron of  the secondary object's orbit around the central star.  The dashed line represents the hydrogen burning limit, and the grey band the deuterium burning limit \citep[$\sim 11-16~{\rm M}_{\rm J}$;][]{Spiegel:2011a}.}
            \label{fig:massRadius}
        \end{centering}
    \end{figure}   

\begin{figure}
        \begin{center}
            \includegraphics[width = 0.5\textwidth, trim = 0cm 0cm 0cm 0cm, clip=false]{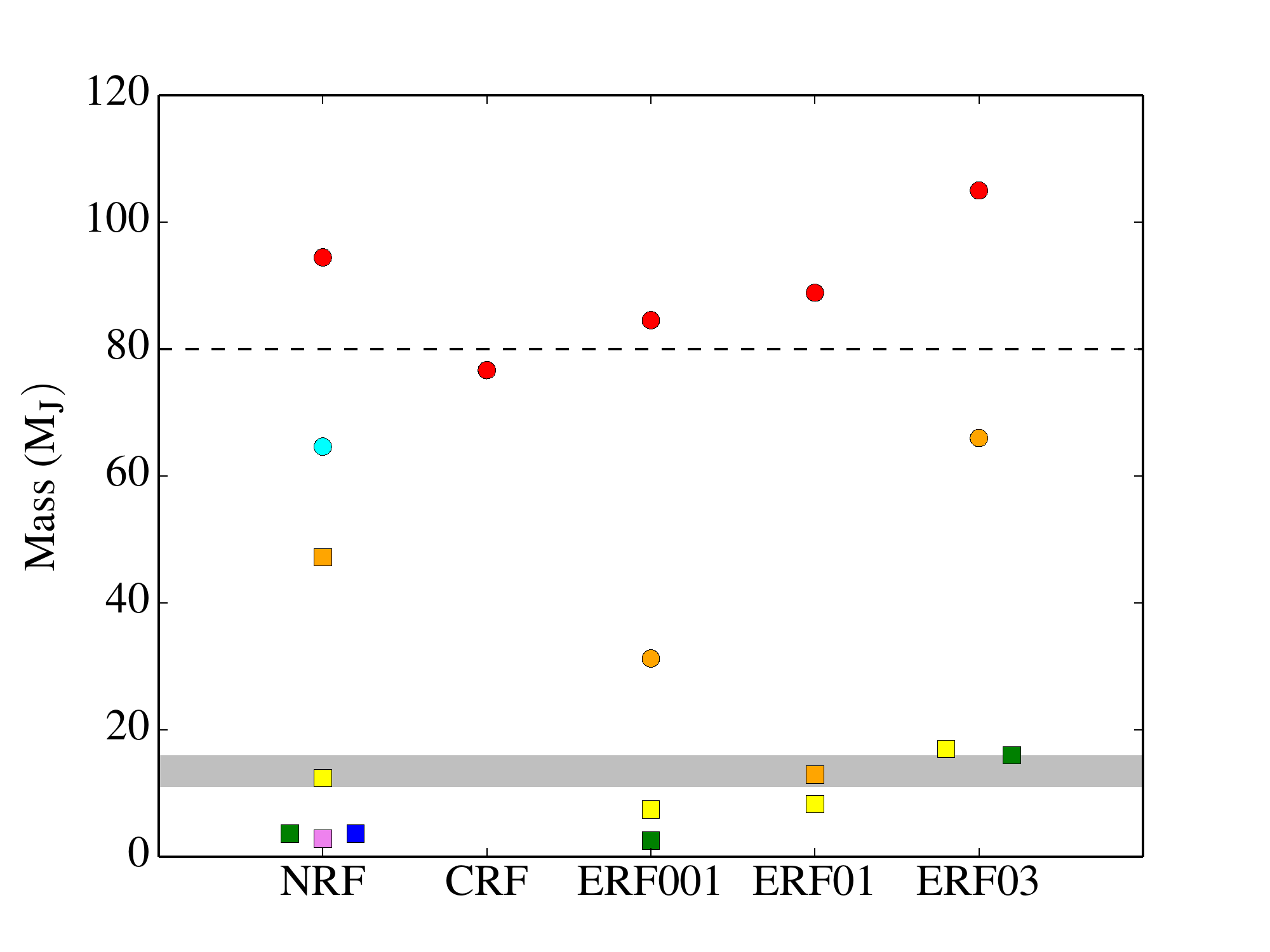}
            \caption{The masses of the secondary objects  at the end of the hydrodynamic simulations (10~kyr). Colour denotes the order in which each secondary object formed; from earliest to latest: red, orange, yellow, green, cyan, blue, violet. Circles and squares correspond to objects that are ultimately bound or ejected (at 200~kyr), respectively. The lower points in the NRF and ERF003 simulations are separated for clarity. The dashed line represents the hydrogen burning limit ($\sim 80~{\rm M}_{\rm J}$). The grey band represents the deuterium burning limit.}
            \label{fig:final_mass}
        \end{center}
    \end{figure}

    In all simulations disc fragmentation occurs beyond radii $\sim 65$ AU (see Figure~\ref{fig:massRadius}a), where the disc is gravitationally unstable and can cool fast enough \citep[e.g.][]{Stamatellos:2009a}. The initial  mass of a fragment is a few $\textup{M}_{\textsc{j}}$, as set by the opacity limit for fragmentation \citep{Low:1976a,Rees:1976a}. The masses of the secondary objects at the end of the hydrodynamical simulations are shown in Figure~\ref{fig:final_mass}. The first object that forms in each simulation generally migrates inwards and accretes enough mass to become a low-mass star; this object remains ultimately bound to the central star. All secondary objects increase in mass as they accrete gas from the disc. However, roughly half of the objects formed in each simulation (excluding continuous radiative feedback) remain as planets by the end of the hydrodynamical simulation as shown in Figure \ref{fig:final_mass}. 

    \begin{figure}
        \begin{center}
            \includegraphics[width = 0.48\textwidth, trim = 0cm 0cm 0cm 0cm, clip=true]{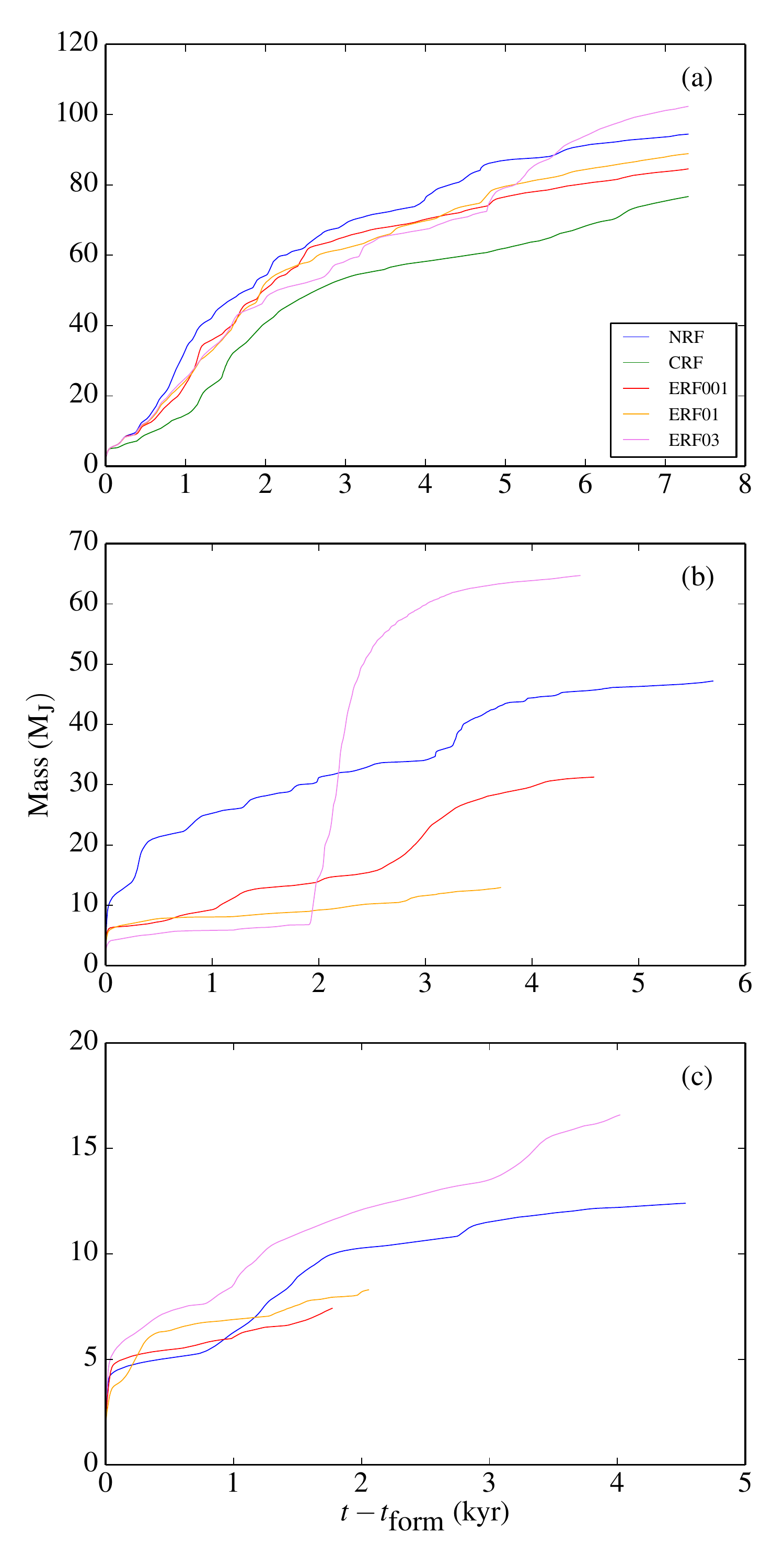}
            \caption{The mass evolution of the first 3 secondary objects that form in each of the 5 simulations (for the simulations with episodic radiative feedback the mass refers to the sink mass, i.e. both the object and the inner accretion disc). Time is given with respect to the formation time of each object. The second object in the ERF03 run (b) undergoes a rapid increase in mass as it migrates into a dense region around the central star.}
            \label{fig:sink_mass}
        \end{center}
    \end{figure}
    \begin{figure}
        \begin{center}
            \includegraphics[width = 0.48\textwidth, trim = 0cm 0cm 0cm 0cm, clip=true]{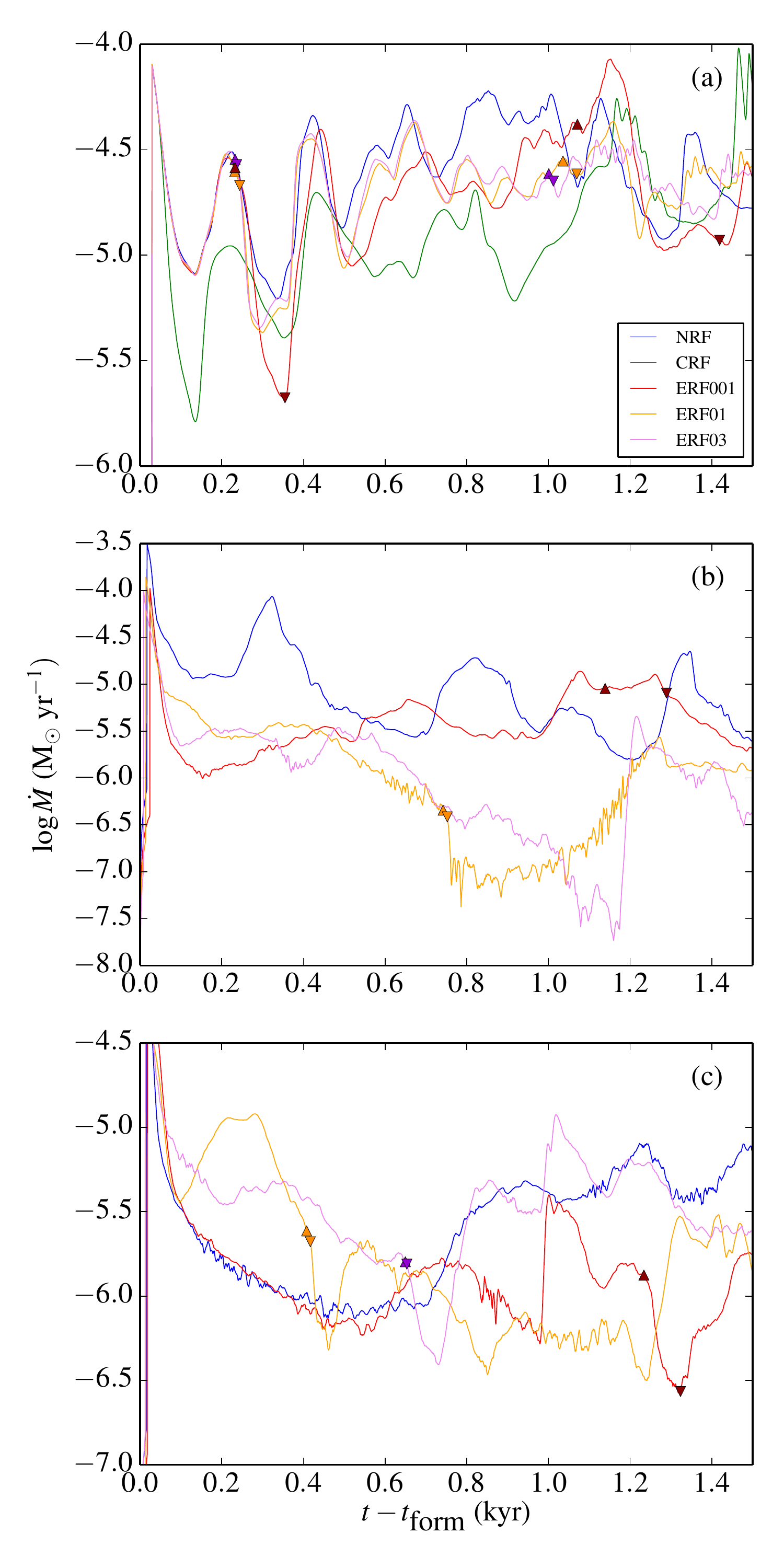}
            \caption{Mass accretion rates onto  the first 3 secondary object that form in each of the 5 simulations (for the simulations with episodic radiative feedback the accretion rates onto the sinks are plotted). Time is given with respect to the formation time of each object. Upward pointing triangles represent the beginning of accretion episodes in the ERF runs. These are followed by corresponding downward pointing triangles denoting the end of episodes.}
            \label{fig:sink_accretion}
        \end{center}
    \end{figure}

    In the continuous radiative feedback simulation (CRF) the mass growth of the secondary object is mildly suppressed (Figures~\ref{fig:sink_mass}a,\ref{fig:sink_accretion}a) due to an increased outward thermal pressure, so that the final mass of the object is within the brown dwarf regime. Secondary objects that form at later times tend to have lower masses (Figure~\ref{fig:final_mass}).

    Episodic feedback also mildly suppresses the mass growth of the first secondary object that forms (Figure~\ref{fig:sink_mass}a). Its effect is more pronounced for the second secondary object (Figure~\ref{fig:sink_mass}b). However, the mass growth of each object also depends on where the object forms in the disc and how it interacts both with other objects and with the spiral structure of the disc. Therefore, the mass growth of an object can be rather erratic, e.g. for the second object at around 2~kyr (Figure~\ref{fig:sink_mass}b). Specifically, this object migrates into the high density region surrounding the central star where it rapidly accretes a large amount of gas  (see Figure~\ref{fig:ea1_03}, 7.2-8 kyr). The effect of episodic accretion is to suppress mass accretion during/after the outburst (e.g. Figure~\ref{fig:sink_accretion}a compare NRF and ERF runs after the first outburst; also seen in Figures~\ref{fig:sink_accretion}b, c). However, the accretion rate  is restored to  to its previous value within $200-400$~yr. Ultimately, there is no strong anti-correlation between the mass that an object and the number and duration of the episodic outbursts it undergoes.

    We find a population of planetary-mass objects on wide orbits  ($100-800$~AU) around the central star. However, these objects are loosely bound to the central star and could be liberated into the field becoming free-floating planets. We follow the evolution of these systems using  NBODY simulations. Indeed we find that all  planetary-mass objects are ejected from the discs (Figure~\ref{fig:final_mass}c); what is left behind is a central star with low-mass star or brown dwarf companions. Consequently, it is unlikely that the observed wide-orbit giant planets \citep{Kraus:2008a, Kraus:2014a, Marois:2008a, Faherty:2009a, Ireland:2011a,Kuzuhara:2011a,Kuzuhara:2013a, Aller:2013a,Bailey:2014a,Rameau:2013b,Naud:2014a,Galicher:2014a,Macintosh:2015a} may form by disc fragmentation, unless somehow the mass growth of secondary objects forming in the disc is suppressed. On the other hand disc fragmentation may readily produce free floating planets and brown dwarfs \citep{Stamatellos:2009a,Hao:2013a, Li:2015b,Vorobyov:2016a}. 

    Note however that in order to follow the long term evolution of the system we have ignored the effect of the gas once the hydrodynamical simulation has evolved for 10~kyr.  The effect of gas is to stabilize the system. Therefore it is possible that some of these planets may remain bound to the central star. However, they should co-exist with a higher mass object (like e.g. a low-mass star or a brown dwarf) and they may accrete enough mass to become brown dwarfs.

    \subsection{Caveats of sink particles} 
    \label{sub:caveats_of_sink_particles}

    	Sink particles are used in the simulations to prevent large running times. In dense regions, timesteps become very short and without sinks the simulation effectively stalls.
	
		In our simulations a sink particle is created when the density exceeds  $10^{-9} \textup{ g cm}^{-3}$. It is therefore assumed that if a proto-fragment reaches this density it will continue to contract  to  heat to $\sim 2000$~K such that  molecular hydrogen dissociates to initiate the second collapse. The proto-fragment will ultimately reach stellar densities  ($\sim 1\textup{ g cm}^{-3}$) to become a bound object. The density threshold used for sink creation is higher than the density required for the formation of the first hydrostatic core ($\sim 10^{-13}\textup{ g cm}^{-3}$). Therefore, the proto-fragment at this stage contracts on a Kelvin-Helmholtz timescale. The time that it takes a proto-fragment to evolve from the first to second hydrostatic core is $\sim 1-10$~kyr \citep{Stamatellos:2009d}. Thus, it is possible that some of the proto-fragments may get disrupted e.g. by interactions with spiral arms and/or tidal forces, and dissolve \citep{Stamatellos:2009d, Zhu:2012a,Tsukamoto:2013a}.
		
		Another limitation in the use of sink particles relates to their size. We assume that the sink radius of secondary objects that form in the disc is 1~AU, which roughly corresponds to  the size of the first hydrostatic core during the collapse of a proto-fragment \citep{Masunaga:2000a, Tomida:2013a, Vaytet:2013a}. The size of the Hill radius of proto-fragments that form in the disc is on the order of a few AU.  Therefore, a significant fraction of the accretion disc around a proto-fragment is not resolved. The flow of material from the sink radius to the secondary object is considered to be instantaneous, whereas, in reality, there is a time delay. This results in increased accretion  onto secondary objects, which in the case of continuous feedback, results in an increased luminosity. As such, we may overestimate the effect of luminosity on disc fragmentation. However, for the episodic accretion runs we employ a sub-grid model (within a sink radius) based on an $\alpha$-viscosity prescription that allows gas to flow (episodically) onto the secondary object (see Section \ref{subsub:radiative_feedback_from_secondary_objects}). Even in this case, the accretion rate is possibly overestimated, as the inner accretion disc within the sink ($<1$~AU) does not exchange angular momentum with the rest of the accretion disc \citep[for an additional discussion of this issue see][]{Hubber:2013b}. Nevertheless, considering the uncertainties in $\alpha_{\textsc{mri}}$ (which in effect modulates the accretion of material onto the secondary objects and for which we examine a wide range of values, all of which lead to similar outcomes) we have confidence that the choice of sink size does not qualitatively affect the results of this paper regarding  the effect of radiative feedback on disc fragmentation.

    
\section{Conclusions} 
\label{sec:conclusions}

    We have performed SPH simulations of gravitationally unstable protostellar discs in order to investigate the effect that radiative feedback from secondary objects formed by fragmentation has on disc evolution. We have considered three cases of radiative feedback from secondary objects: {\bf (i)} No radiative feedback: where no energy from secondary objects  is fed back into the disc. {\bf(ii)} Continuous radiative feedback: where energy, produced by accretion of material onto the surface of the object is continuously fed back into the disc. {\bf (iii)} Episodic radiative feedback: where accretion of gas onto secondary objects is episodic, resulting in episodic radiative feedback. Our findings are summarised as follows:
    
    \begin{itemize}
        \item Radiative feedback from secondary objects that form through gravitational fragmentation stabilises the disc, reducing the likelihood of subsequent fragmentation. When there is no  radiative feedback from secondary objects, 7 objects form, compared to a single object forming when radiative feedback is continuous. When  radiative feedback happens in episodic outbursts, $3-4$ objects form. This is because the disc cools sufficiently to become gravitationally unstable between the outbursts. All objects in the three different radiative feedback cases that we examine here form at radii $> 65$~AU, with initial masses of a few $\textup{M}_{\textup{J}}$. \newline

        \item The mass growth of secondary objects is mildly suppressed due to their radiative feedback. The mass of the first object that forms within the disc is generally larger when there is no radiative feedback; in the case when radiative feedback is continuous the mass of the first secondary object is the lowest. Episodic radiative feedback tends to reduce the mass accretion rate onto a secondary object during and after an episode outburst. However, the accretion rate is restored to its previous value relatively quickly (within $\sim200 - 400$~yr). \newline

        \item The intensity and the duration of an outburst (which in our models is determined by the effective viscosity due to the magnetorotational instability, $\alpha_{\textsc{mri}}$) does not affect  the number of objects that form within the disc when episodic radiative feedback is considered. The total duration of the radiative feedback outbursts is not long enough to  fully suppress disc fragmentation. However, we find that $\alpha_{\textsc{mri}}$ affects the average mass of the objects formed: lower $\alpha_{\textsc{mri}}$ results into lower mass secondary objects. Moreover, subsequent fragmentation happens faster  for higher  $\alpha_{\textsc{mri}}$, as the first outburst finishes faster. The first object that forms in each case undergoes a larger inward migration for increased values of $\alpha_{\textsc{mri}}$. \newline

        \item Regardless of the type of radiative feedback, we find that the first object that forms within the disc, remains ultimately bound to the central star. It accretes mass while it generally migrates inwards. Brown dwarfs also form in the simulations and a fraction of them remain bound to the central star. Gravitational fragmentation may therefore provide a method for the formation of intermediate separation, low-mass-ratio binary systems. \newline
        
        \item A significant fraction ($\sim40$\%, dropping to $\sim20$\% if the estimated final mass is considered) of the secondary objects formed by disc fragmentation are planets, regardless of the type of radiative feedback. However, every planet that forms within the disc is ultimately ejected from the system.   We do not find any giant planets that remain on wide-orbits around the central star. Secondary objects that form and remain within the disc accrete enough mass to become brown dwarfs, even in the case where radiative feedback suppresses gas accretion. Thus, gravitational fragmentation may produce free-floating planets and brown dwarfs, but not wide-orbit gas giant planets, unless the mass growth of fragments forming in a young protostellar disc is further suppressed. 

\end{itemize}


\section*{Acknowledgements} 
\label{sec:acknowledgements}

    We would like to thank the referee A. Boley of his thorough review of the paper. The simulations presented in this paper were performed at UCLan High Performance Computing Facility {\sc wildcat}. Surface density plots were produced using the \textsc{splash} software package \citep{Price:2007b}. N-body simulations were performed by a code originally developed by D.~Hubber, who we thank for the support provided. AM is supported by STFC grant ST/N504014/1. DS is partly supported by STFC grant ST/M000877/1.
    


\bibliography{bibliography}{}
\bibliographystyle{mnras}

\bsp	
\label{lastpage}
\end{document}